\documentclass[letterpaper,journal]{IEEEtran}
\usepackage{amsmath,amsfonts}
\usepackage{algorithmic}
\usepackage{algorithm}
\usepackage{array}
\usepackage[caption=false,font=normalsize,labelfont=sf,textfont=sf]{subfig}
\usepackage{textcomp}
\usepackage{setspace}

\usepackage{stfloats}
\usepackage{url}
\usepackage{verbatim}
\usepackage{graphicx}
\usepackage{cite}
\usepackage{mathtools}
\usepackage{cleveref}

\usepackage{gensymb}
\usepackage{multirow}
\usepackage{eurosym}
\usepackage[colorinlistoftodos]{todonotes}
\usepackage{booktabs} 
\usepackage{capt-of}
\usepackage{tablefootnote}
\usepackage{makecell}
\usepackage{xcolor}

\makeatletter
\newcommand{\specialcell}[1]{\ifmeasuring@#1\else\omit$\displaystyle#1$\ignorespaces\fi}
\makeatother

\begin{document}
	\title{Optimization-based framework for low-voltage grid reinforcement assessment under various levels of flexibility and coordination}
	
	\author{\IEEEauthorblockN{Soner Candas$^{1}$\IEEEauthorrefmark{1},  Beneharo Reveron Baecker$^{1}$, Anurag Mohapatra$^{2}$, Thomas Hamacher$^{1}$\\} 
		\IEEEauthorblockA{$^{1}$Chair of Renewable and Sustainable Energy Systems, Technical University Munich
			\\
            $^2$Center for Combined Smart Energy Systems, Technical University of Munich \\
			\IEEEauthorrefmark{1} Corresponding author. Address: Lichtenbergstrasse 4a, 85748 Garching, Germany. E-Mail: soner.candas@tum.de}

            }
	
	
	\markboth{Manuscript accepted on April 15th, 2023 for the Elsevier Applied Energy journal (Candas et al.)}%
	{Shell \MakeLowercase{\textit{et al.}}: A Sample Article Using IEEEtran.cls for IEEE Journals}

	\maketitle
	
	\begin{abstract}
		The rapid electrification of residential heating and mobility sectors is expected to drive the existing distribution grid assets beyond their planned operating conditions. This change will also reveal new potentials through sector coupling, flexibilities, and the local exchange of decentralized generation. This paper thus presents an optimization framework for multi-modal energy systems at the low voltage (LV) distribution grid level. In this, we focus on the reinforcement requirements of the grid and the techno-economic assessment of flexibility components and coordination between agents. By employing a multi-level approach, computational complexity is reduced, and various levels of coordination and flexibilities are implemented. We conclude the work with a case study for a representative rural grid in Germany, in which we observe high economic potential in the flexible operation of buildings, majorly thanks to better integration of photovoltaics. Across all paradigms barring a best-case benchmark, grid reinforcement based on a mix of passive and active measures was necessary. A synergy effect is observed between flexibilities and coordination, as their combination reduces the peaks to the extent of completely avoiding grid reinforcement. The presented framework can be applied with a wide range of grid and component types to outline the broad landscape of future LV distribution grids.

	\end{abstract}
	
	\section{Introduction}
	 Climate change mitigation and geopolitical influences on exploding energy prices are two of the current biggest challenges for society at large. A transition away from fossil fuels, through renewable, distributed energy resources (DERs) such as solar photovoltaics (PV) and wind energy, would ensure emission reductions in line with climate change policy goals and greater price stability. Not only are these measures essential for a sustainable energy future, but also motivated by the increasingly price-competitive technologies \cite{GovindaR.Timilsina.2021}.
The decarbonization plans must also involve heating and mobility sectors through electrification strategies. Electrification leverages the advantages of high-efficiency technologies such as residential heat pumps (HPs) and battery electric vehicles (BEV). Correspondingly, the German Network Development Plan considers a growth scenario for HPs from 1.2 to up to 14.3 million units, and BEVs from 1.2 million to up to 31.7 million for the 2021 to 2037 period \cite{GermanFederalNetworkAgency.07.2022}.

Yet, conventional planning of the low-voltage (LV) distribution grids, despite being over-dimensioned to some extent, had not anticipated a scale of electrification that is in line with the current decarbonization goals. New planning constraints have also emerged from the sector-specific peculiarities of demand. Heating systems have high simultaneity of loads through their correlation to lower ambient temperatures \cite{JennyLove.2017}, which will cause significant increases in peak electrical demands. Electrified mobility systems will have to navigate around coincident arrival-to-home behavior and, thus, limited and similar charging hours \cite{AgoraVerkehrswende.}, \cite{GermanFederalNetworkAgency.2020}. 

There exists a general consensus that the large-scale adoption of these technologies at the household level will cause significant reinforcement requirements in the LV distribution grids \cite{Eurelectric.January2021}. On the other hand, there is also an opportunity emerging in this trend that could mitigate the extent of the grid fortification required to feasibly transform all the energy sectors. Sectoral interactions, load shifting, and quick dispatch properties of DERs offer new flexibility capabilities. However, a holistic coordination paradigm spanning the entire LV grid is required to operate the components in individual households with a grid-relieving agenda and aggregate the flexibilities.

	\subsection{Literature review}
	A vast body of literature tackles the topic of distribution grid reinforcement requirements in the wake of electrified heating and mobility demands, as listed in the state of research provided by Thormann et al. \cite{BerndThormann.2020}. These findings motivate the development of systematic methods to solve the non-convex optimization problem at the heart of distribution grid planning and operation. 
    
    A multitude of possible grid-relieving measures exist, including but not limited to voltage regulators, reactive power compensation devices, parallel cable reinforcement, feed-in management, and grid re-configuration \cite{BenjaminBayer.2018}. These are distinct measures, all of which, in principle, support the grid state against violating voltage and loading restrictions. Due to the high variety of options, combined with the non-linear properties and the complex structure of the electrical grid, a systematic planning routine that identifies the optimal combination of measures poses a highly-complex mathematical challenge. The presence of DERs brings new, less predictable dynamics to the grid operation, adding to this complexity \cite{SergeyKlyapovskiy.2020}.
 
    Sedghi et al. \cite{MahdiSedghi.2016} present a comprehensive review of various heuristic algorithms applied to optimize distribution grid planning with new sectors integrated as electrical loads. Specifically, the works of Saboori et al. \cite{HedayatSaboori.2015} and Koopmann et al. \cite{SimonKoopmann.10620131092013,SimonKoopmann.2013} demonstrate the potential of particle swarm optimization and genetic algorithm for the distribution grid expansion problem. Bakken et al. \cite{Bakken.2020} develop a grid planning algorithm based on discrete alternatives, motivated by the overvoltage instances caused by growing shares of PV systems in LV grids.

	Correspondingly, extensive literature also exists on the convex optimization or mixed-integer linear programming (MILP)-based solution to the grid reinforcement problem. These formulations require certain simplifications for the grid representation in the model. Nevertheless, they have the advantage of the availability of efficient solvers. This brings forth i) the ease of reaching the global optimum while ii) utilizing a broader degree of freedom in the system, incorporating the investment and operation decisions for additional system assets not limited to the grid. The utility of such convex formulations is presented, for instance, in the work of Lopez et al. \cite{JulioLopez.2020}.

	The co-optimization of the multi-modal, non-grid assets also facilitates the analysis of complex sector coupling interactions. These include, but are not limited to, i) trade-offs between the electrification of household assets and the ensuing need for grid expansion, and ii) the potential in the flexible operation of household assets for mitigating grid reinforcement requirements. Pursuing the idea of the holistic optimization of the supply region, Morvaj et al. \cite{BoranMorvaj.2016} develop a MILP-based framework assisted with a genetic algorithm. They investigated the coupled electricity and heat supply at the district level and analyzed the influence of distributed energy sources on the necessary grid reinforcement. An expansion of this to include district heating is presented in \cite{BoranMorvaj.2017}. Mashayekh et al. \cite{SalmanMashayekh.2017} build on this line of work, including cooling demands and minimizing network losses. 
	
	A set of literature focuses primarily on the potential of various flexibility measures to avoid extensive grid reinforcement by exploring solutions in demand-side management for peak shaving and the storage of electricity and heat. Spiliotis et al. \cite{KonstantinosSpiliotis.2016} investigate the potential of demand-side flexibility in avoiding distribution grid expansions by developing a model that represents a market mechanism. Resch et al. \cite{MatthiasResch.2021} compare various pre-defined control mechanisms for flexibility supply through PV and batteries by employing them in grid simulation. These studies give a thorough insight into the benefits of the flexibilities provided by non-grid assets. However, interactions between the capacities of the grid and flexibility components are not comprehensively understood, as system-wide optimization is lacking.
	\newpage
	\subsection{Contribution of this work}	
	Building on the above-mentioned literature, this work introduces an optimization-based framework using a mixed-integer linear programming approach with a composite grid reinforcement formulation. This structure allows us to explore a near-global optimum for the whole system that can be achieved through a combined optimization of LV distribution grid reinforcement and flexibility measures. In this process, each building is defined as a so-called "energy hub" \cite{MartinGeidl.2007}, where the flexible coverage of all energetic demands, i.e., electricity, heating, and mobility, along with the interactions in between, are considered. As for the grid planning, a mixed-integer formulation is developed that facilitates the co-optimization of 1) passive reinforcement measures such as discrete transformer and cable replacement and 2) active measures such as remote PV feed-in curtailment. This way, the cost-optimal mix of the various measures available to the grid planner can be determined. This approach is eminently tractable, as demonstrated by a case study for the complete electrification of a representative German LV distribution system with adequate temporal resolution. Moreover, we validate the accuracy of the electrical flows and voltages with a post-optimization non-convex power flow simulation step, not only ensuring physically feasible planning but also quantifying the network losses in detail.
	
    In this process, one recognizes that the optimum achieved through such a MILP approach posits a single social planner minimizing a single cost function for the whole system. Accordingly, it assumes the full cooperation between the active agents in the system, i.e., the households and the distribution grid planner. Such system planning and operation is far from practice in the current regulatory framework. Therefore, we put these results into perspective by implementing more realistic, albeit sub-optimal, planning schemes. Through such comparison, we attempt to quantify the value of the flexibility components and inter-agent cooperation in future LV distribution grids. While similar analyses have been made in the work of Grimm et al. \cite{VeronikaGrimm.2020}, this study extends the analysis to heating and mobility demands with endogenous investments of all assets. To the best of the authors' knowledge, this framework is the first that combines i) a multi-modal system model at the distribution grid level, ii) investigating the requirements of various grid reinforcement measures iii) under various flexibility and coordination levels. 
     
    	\begin{figure*}[!t]
		{\centering
		\includegraphics[width=1\linewidth]{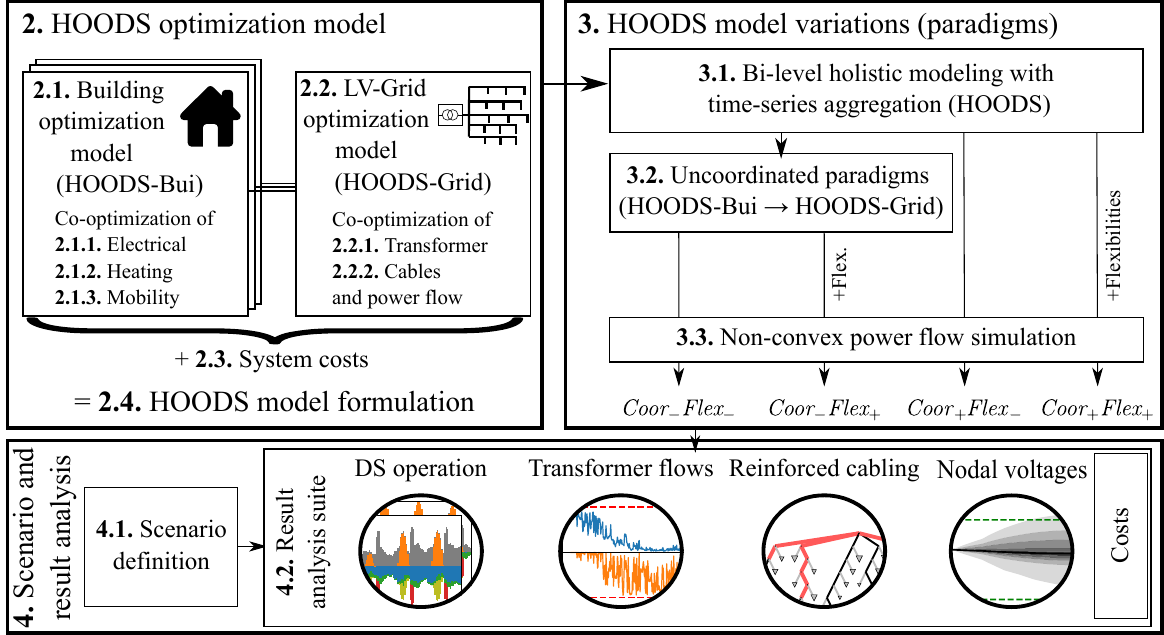}
		\caption{Structure of the proposed optimization-based framework for the assessment of LV-grid reinforcement requirements under various flexibility and coordination settings. The numbering represents the relevant sections in the paper where each module of the framework is discussed.}
		\label{fig:concept}}
	\end{figure*}

    The workflow of the proposed optimization framework, along with the section numbers where each module is described, is illustrated in Figure \ref{fig:concept}. The remainder of the work is structured in the following way. The holistic optimization of distribution system (HOODS) model formulation and the implementation of the planning paradigms based on it are described in Sections \ref{sec: model} and \ref{sec: paradigms}, respectively. In Section \ref{sec: casestudy}, the data set forming the case study is described along with the discussion of the case study results. Section \ref{sec:conclusion} concludes the work with a summary of the findings and outlook on future research.   
    \subsection{Definitions}
    In this work, the following definitions are adopted:
    \begin{itemize}
    \item \textit{Building components}: Technical components located in buildings, which serve for the generation, storage and conversion of energy (see Figure \ref{fig:res}).
    \item \textit{Buildings} or \textit{households}: Agents which demand energy in various forms and operate their building components accordingly. They also have control over investment decisions over the building components. 
    \item \textit{Low-voltage (LV) distribution grid}: The part of the electrical power grid where the electrical energy usually reaches residential consumers. European LV grids usually have a three-phase rated voltage of 400V. 
    \item \textit{Grid components}: Technical components that constitute the electrical network, which transports active and reactive power across the grid. In this work, they consist of LV transformers (fixed-ratio or those with an on-load tap changer), distribution cables, and a reactive power compensator.
    \item \textit{Grid planner}, \textit{grid operator} or \textit{distribution grid operator (DSO)}: Agent responsible for the coverage of the households' electrical demand without violating the network restrictions. They also decide upon the reinforcement of the grid components to maintain a safe network operation. 
    \item \textit{Distribution system (DS)}: The totality of agents involved in the energy supply in the LV grid level. The DS consists of all buildings in the LV grid and the grid planner. 
    \item \textit{Model}: Mathematical representation of the cost-optimal operation and planning of the DS, which can be formulated as an \textit{optimization problem} that can be solved by an optimization \textit{solver}.
    \item \textit{Model variation} or \textit{paradigm}: A specific variation on the model, which is characterized by a different degree of coordination and flexibility in the DS planning and operation.
    \end{itemize}
    
	\section{Holistic optimization of distribution system (HOODS)}\label{sec: model}
	This section elaborates on the mathematical formulation of the holistic optimization of distribution system (HOODS) model, which is the central element of the proposed framework. The model i) deals with a holistic optimization of the building and grid components, ii) allows simultaneous formulations for the expansion and the yearly operation while iii) keeping a tractable MILP formulation. The corresponding optimization problem is formulated and implemented by extending the open-source energy system modeling framework \textit{urbs}\footnote{GitHub repository: \url{https://github.com/tum-ens/urbs}.}. Among other applications, a version of the \textit{urbs} framework has been previously used to investigate transmission-distribution system interactions in a national energy system \cite{BeneharoReveronBaecker.2022}. 	
	The following subsections will discuss the modeling approaches taken for the building and grid components, followed by the representations of the system costs, and the resulting HOODS problem formulation. 
	\subsection{Modeling of building components}\label{sec: bui_comp}
	\par
\begin{figure*}[!htb]
		\centering
		\includegraphics[width=0.9\linewidth]{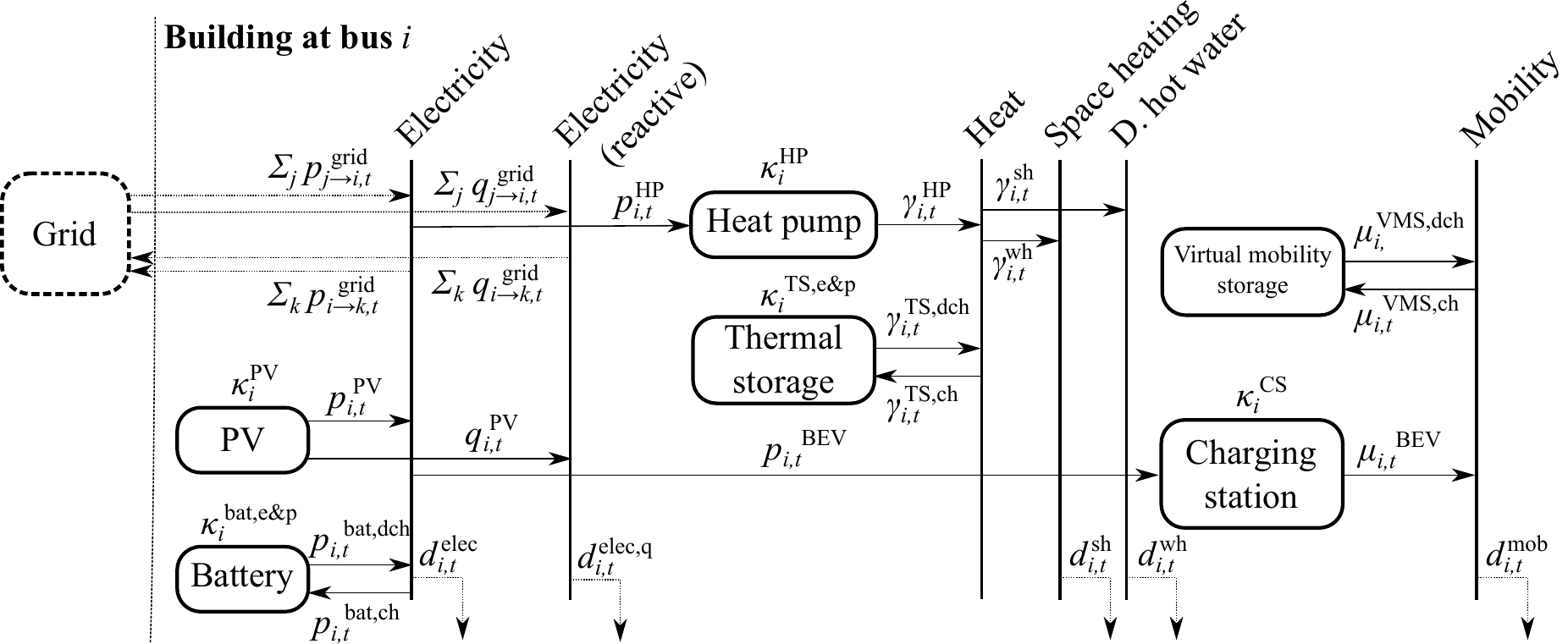}
		\caption{Sector-coupled reference energy system for a fully-electrified building along with the corresponding energetic flows.}
		\label{fig:res}
	\end{figure*}

 Buildings can be located at a subset $\mathcal{I}^b$ of the given set of network nodes $\mathcal{I}$ within the LV distribution grid (building $i \in \mathcal{I}^b\subseteq\mathcal{I}$). For each building, respective electricity, space heating, domestic hot water, and mobility demands $d_{i,t}^{\{...\}}$ have to be covered for each time step $t\in\mathcal T$, specifically on an hourly basis. If no building is located in a given network bus, $d_{i,t}^{\{...\}}=0$. In a scenario of complete electrification, these demands can be covered by the set of technologies as depicted in Table \ref{tab:techopts}. The corresponding reference energy system in Figure \ref{fig:res} illustrates the energetic interactions within each building.
	
	\begin{table}[!htb]\footnotesize
		\caption{Building components defined for each energy demand.\label{tab:techopts}}
		\centering
		\begin{tabular}{c|c|c}
			
			Energy demand & Supply & Flexibility\\
			\hline
			Electricity & Solar PV, grid electricity & Battery storage\\
			Heat (space\&water) & Air source heat pump & Heat storage\\
			Mobility & Charging station (11 kW) & Overnight charging\\		
			
		\end{tabular}
	\end{table}
	\subsubsection{Electrical components} Building components dealing with local generation and storage of electricity consist of PV units and household batteries, respectively. A linear modeling of the PV operation for a given building at $i$ and an hour $t\in\mathcal{T}$ is achieved via the following constraints:
	\begin{subequations}
		\begin{align}
		0 &\leq p_{i,t}^\text{PV} = \kappa_{i}^\text{PV}\cdot cf_t^\text{PV} - {p}_{i,t}^\text{PVcurt}, \label{eqn:pv}\\
		-\tan\left({\phi_\text{min}^\text{PV}}\right)\cdot p_{i,t}^\text{PV} &\leq q_{i,t}^\text{PV} \leq \tan\left({\phi_\text{min}^\text{PV}}\right)\cdot p_{i,t}^\text{PV} , \label{eqn:pq}	
		\end{align}
	\end{subequations}
	where $p_{i,t}^\text{PV}$ and $q_{i,t}^\text{PV}$ are the hourly active and reactive power generation from PV respectively, ${p}_{i,t}^\text{PVcurt} \geq 0$ is the curtailed PV generation, $\kappa_{i}^\text{PV}$ is the endogenous installed capacity of PV, $cf_t^\text{PV}$ is the capacity factor (normalized possible generation) at the given hour and $\cos\phi_\text{min}^\text{PV}$ is the minimum power factor that the PV unit is allowed to operate at\footnote{For instance, the German VDE-AR-N 4105 norm sets a mandatory minimum power factor of up to $\cos{\phi_\text{min}^\text{PV}} = 0.95$ for units between 3.68 kVA and 13.8 kVA, and $\cos{\phi_\text{min}^\text{PV}} = 0.9$ for a module capacity larger than 13.8 kVA.}. The active and reactive power supply is decoupled---the reactive power provision from PV at a given time is assumed not to reduce the active power capability. Moreover, it is assumed that the PV inverter generating the reactive power has adequate capacity to accommodate the additional apparent power. 
	
	The capacity expansion of PV is restricted by the building rooftop area $A_i$, with a roof area usage factor of 0.1 kW per m$^2$:
	\begin{equation}
	0 \leq \kappa_i^\text{PV} \leq \left(0.1 \text{ kW/m}^2 \right) \cdot A_i \cdot \beta_i^\text{PV},\label{eq: pvlimit}
	\end{equation} 
	where $\beta_i^\text{PV}\in\{0,1\}$ is the binary variable representing the decision to install a PV module on the rooftop of a building. 
	
	Overproduction of PV electricity at each house can be integrated via the intelligent operation of household batteries. The \crefrange{eqn:bat_bal}{eqn:bat_e2p} govern the operation of a battery:
	\begin{subequations}
		\begin{align}
		\epsilon_{i,t}^\text{bat,e} &= \epsilon_{i,t-1}^\text{bat,e}\cdot (1-\delta^\text{bat}) + \eta^\text{bat,ch} \cdot p_{i,t}^\text{bat,ch} - \dfrac{p_{i,t}^\text{bat,dch}}{\eta^\text{bat,dch}}\label{eqn:bat_bal}, \\
		0&\leq p_{i,t}^\text{bat,dch} \leq \kappa_i^\text{bat,p}\label{eqn:bat_cap_dch},\\
		0&\leq p_{i,t}^\text{bat,ch} \leq \kappa_i^\text{bat,p} \label{eqn:bat_cap_ch},\\
		0&\leq \epsilon_{i,t}^\text{bat,e} \leq \kappa_i^\text{bat,e}\label{eqn:bat_cap_e},\\
		\epsilon_{i,0}^\text{bat,e} &= \epsilon_{i,T_\text{end}}^\text{bat,e},\label{eqn:bat_cyc}\\
		\kappa_i^\text{bat,e} &= \kappa_i^\text{bat,p} \cdot etp^\text{bat},\label{eqn:bat_e2p}
		\end{align}
	\end{subequations}
	where the variable $\epsilon_{i,t}^\text{bat,e}$ expresses the energy content of the battery unit at a given time, $p_{i,t}^\text{bat,ch}, p_{i,t}^\text{bat,dch}$ are the hourly charged and discharged amounts of active power, and $\kappa_{i}^\text{bat,p},\kappa_{i}^\text{bat,e}$ are the power and energy capacities of the battery unit. The constraint (\ref{eqn:bat_bal}) governs the state-of-charge of the battery subject to the charging and discharging efficiencies of $\eta^\text{bat,ch},\eta^\text{bat,dch}$ and the hourly relative self-discharge rate $\delta^\text{bat}$. The \crefrange{eqn:bat_cap_dch}{eqn:bat_cap_e} restrict the battery operation and the state-of-charge with their respective capacities, the equation (\ref{eqn:bat_cyc}) enforces the cyclic operation of the battery, and the equation (\ref{eqn:bat_e2p}) ties the power and energy capacities of the battery by a preset energy-to-power ratio of $etp^\text{bat}$. No reactive power capabilities are defined for the battery inverters as there are no current norms prescribing such a requirement. Operational nonlinearities pertaining to current, voltage, or cell aging are neglected in the linear modeling of the household battery at the energy level. Due to its computational advantages, such "generic" storage modeling has been common practice, as also adopted by a sheer majority of energy system modeling frameworks \cite{Groissbock.2019}.
	
	At each timestep $t\in\mathcal{T}$, the energy balance for active and reactive power is maintained by the following constraints for a given node $i\in \mathcal I$:
	
	\begin{subequations}
		\begin{align}
		\begin{multlined}[t]
		p_{i,t}^\text{PV} + p_{i,t}^\text{bat,dch} + \sum_{\mathclap{j \in N^{-}(i) }}p_{j\rightarrow i,t}^\text{grid} \\= d_{i,t}^\text{elec} + p_{i,t}^\text{bat,ch} + p_{i,t}^\text{BEV}+ p_{i,t}^\text{HP} + \sum_{\mathclap{k \in N^+(i) }}p_{i\rightarrow k,t}^\text{grid},\end{multlined} \label{eqn:pbal}\\
		q_{i,t}^\text{PV} + \sum_{\mathclap{j \in N^{-}(i) }}q_{j\rightarrow i,t}^\text{grid} = d_{i,t}^\text{elec,q}  + \sum_{\mathclap{k \in N^+(i) }}q_{i\rightarrow k,t}^\text{grid},\label{eqn:qbal}
		\end{align}	
	\end{subequations}
	
    where $\sum_{j \in N^{-}(i) }p,q_{j\rightarrow i,t}^\text{grid}, \sum_{k \in N^+(i) }p,q_{i\rightarrow k,t}^\text{grid}$ are the in- and outflows of active and reactive power from all predecessor and successor nodes $j \in N^{-}(i)$ and $k \in N^{+}(i)$ respectively\footnote{For a radial network configuration typical for low-voltage distribution grids, the set of predecessor nodes has a cardinality of one ($|N^{-}(i)|=1\; \forall i \in \mathcal{I}$).}, $ p_{i,t}^\text{BEV}, p_{i,t}^\text{HP}$ are the active power feed-in to the BEV and to the heat pump respectively, and $d_{i,t}^\text{elec}, d_{i,t}^\text{elec,q}$ are the exogenously provided time series for the active and reactive power demand of the electrical appliances in the building. In the general sense, the electrical gain terms on the left-hand side are balanced by the loss terms on the right-hand side. For nodes where no buildings are located ($i\in \mathcal I\setminus\mathcal I^b$), the equations (\ref{eqn:pbal})--(\ref{eqn:qbal}) simply reduce to the balance of in- and outflows. 
	\subsubsection{Heat components}
	In our model, the total electrification of building space and domestic hot water demand is provided by heat pumps (HPs). In particular, air source HPs have been selected over ground- and gas source  variants. Despite having efficiency disadvantages in very low ambient temperatures, air source HPs are the most used type in Europe \cite{PaulaCarroll.2020}. They can be employed independently of the access to groundwater and have significantly lower installation costs than the ground source HPs. Moreover, they may have a low-emission operation as the electricity generation gets decarbonized in the future, unlike the gas source HPs. The \crefrange{eqn:hpcop}{eqn:hpcap} define the power consumption and the capacity restriction of the heat pump:
	\begin{subequations}
		\begin{align}
		\gamma_{i,t}^\text{HP} &= p_{i,t}^\text{HP}\cdot cop_t^\text{HP}, \label{eqn:hpcop}\\
		0\leq p_{i,t}^\text{HP} &\leq \kappa_i^\text{HP},\label{eqn:hpcap}
		\end{align}
	\end{subequations}
	where $\gamma_{i,t}^\text{HP}$ is the heat output from the heat pump, $cop_t^\text{HP}$ is the exogenous coefficient of performance (COP), and $\kappa_i^\text{HP}$ is the electrical capacity of the heat pump. Distinct values for each hour are used to factor in the COP variations depending on the ambient temperature. Heat pumps are also assumed to operate at a temperature higher than 60$\degree$C, which is recommended for domestic hot water against legionella formation. A temperature gradient-dependent coefficient of performance (COP) time series has been incorporated from \cite{HansMartinHenning.2013}. Furthermore, a continuous operation of the heat pump between zero and the full load is allowed in the model, while most commercial single-stage heat pumps may only operate at full load in their current state of maturity. Although this representation is most suitable for the new generation of inverter-driven heat pumps with flexible operation ranges, this is also a valid formulation for single-stage units. The modeled heat pump can operate at full load for only a fraction of the hour, followed by a shutdown for the remaining period, thereby meeting a range of total heat production $\gamma_{i,t}^\text{HP}$ within an hour. 
	Similar to PV, a decision variable for heat pump installations $\beta_i^\text{HP}$ is modeled using the following equation:	
	\begin{equation}
	0 \leq \kappa_i^\text{HP} \leq M \cdot \beta_i^\text{HP},\label{eq: hplimit}
	\end{equation} 
	where $M$ is a large parameter. 
	
	Flexibility in the heat supply is achieved through sensible thermal storage typical to domestic use. Analogous to the battery model (\crefrange{eqn:bat_bal}{eqn:bat_cyc}), the thermal storage is modelled as follows:
	\begin{subequations}
		\begin{align}
		\gamma_{i,t}^\text{TS,e} &= \gamma_{i,t-1}^\text{TS,e}\cdot (1-\delta^\text{TS}) + \eta^\text{TS,ch} \cdot \gamma_{i,t}^\text{TS,ch} - \dfrac{\gamma_{i,t}^\text{TS,dch}}{\eta^\text{TS,dch}}, \label{eqn:ts_bal}\\
		0&\leq \gamma_{i,t}^\text{TS,dch} \leq \kappa_i^\text{TS,p},\\
		0&\leq \gamma_{i,t}^\text{TS,ch} \leq \kappa_i^\text{TS,p}, \\
		0&\leq \gamma_{i,t}^\text{TS,e} \leq \kappa_i^\text{TS,e},\label{eqn:ts_cap_e}\\
		\gamma_{i,0}^\text{TS,e} &= \gamma_{i,T_\text{end}}^\text{TS,e}.\label{eqn:ts_cyc}
		\end{align}
	\end{subequations}
	Coverage of the exogenous hourly space and water heating demands $d_{i,t}^\text{sh} + d_{i,t}^\text{sh}$ is maintained by the energetic heat balance equation (\ref{eqn:heat_balance}):
	\begin{equation}
	\gamma_{i,t}^\text{HP} + \gamma_{i,t}^\text{TS,dch} = \gamma_{i,t}^\text{TS,ch} + d_{i,t}^\text{sh} + d_{i,t}^\text{wh} \label{eqn:heat_balance}.
	\end{equation}
	\subsubsection{Mobility modeling} 
\begin{figure}
		\includegraphics[width=0.8\linewidth]{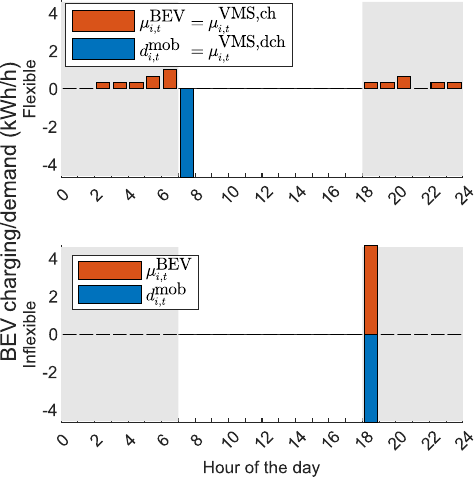}
		\caption{Example daily BEV charging patterns achieved through the constraints (\ref{eqn:bevconv}--\ref{eqn:mob_dem2}).}\label{fig:bevmodes}
    \end{figure}

Additional to the conventional electrical appliances and heating, daily electrified private mobility demands $d_{i}^\text{mob,daily}$ are defined for each building. Two modes of charging can be modeled to cover this demand: i) inflexible charging in the afternoon and ii) flexible overnight charging. In the first mode, the conventional charging behavior is simulated so that the BEV is charged as soon as it reaches home in the late afternoon (e.g., 6 PM). The second mode, in contrast, allows a more flexible, flattened charging profile, distributed between the late afternoon and early morning hours (e.g., 6 PM-7 AM) to avoid simultaneous peaks (see Figure \ref{fig:bevmodes}). 
 
	The charging station operates with the efficiency $\eta^\text{CS}$ and has a capacity of $\kappa_i^\text{CS}$:
	\begin{subequations}
		\begin{align}
		&\mu_{i,t}^\text{BEV} = \eta^\text{CS}\cdot p_{i,t}^\text{BEV}, \label{eqn:bevconv}\\
		0\leq &\mu_{i,t}^\text{BEV}\leq \begin{cases}
		\kappa_{i}^\text{CS}  &\forall t \in \{\text{18:00-7:00}\}\\
		0  &\forall t \in \{\mathcal{T}\setminus\text{18:00-7:00}\}\end{cases}\label{eqn:bevcap}
		\end{align}
	\end{subequations}
    where $\mu_{i,t}^\text{BEV}$ is the energy transferred to the BEV at a given time. Note that the charging is allowed only in a certain period daily (e.g., between 6 PM and 7 PM), as the BEVs are assumed to be not present in their respective building otherwise, and alternative charging possibilities are excluded.
	\newpage
	In order to allow flexible charging, the charging and discharging variables ($\mu_{i,t}^\text{VMS,ch}, \mu_{i,t}^\text{VMS,dch}$) of a "virtual" mobility storage (VMS) are incorporated into the mobility energy balance:
	\begin{align}
	\mu_{i,t}^\text{BEV} + \mu_{i,t}^\text{VMS,dch} &= d_{i,t}^\text{mob} + \mu_{i,t}^\text{VMS,ch}.\label{eqn:mobbal}
	\end{align}
	The VMS is modeled as a lossless storage unit with no self-discharge and its maximum rate of "operation" is restricted by the charging station capacity $\kappa_i^\text{CS}$:
	\begin{subequations}
		\begin{align}
		\mu_{i,t}^\text{VMS,e} &= \mu_{i,t-1}^\text{VMS,e} + \mu_{i,t}^\text{VMS,ch} - \mu_{i,t}^\text{VMS,dch}, \label{eqn:vms_bal} \\
		0&\leq \mu_{i,t}^\text{VMS,dch} \leq \kappa_i^\text{CS},\\
		0&\leq \mu_{i,t}^\text{VMS,ch} \leq \kappa_i^\text{CS}, \label{eqn:vms_cap_ch}\\
		\mu_{i,0}^\text{VMS,e} &= \mu_{i,T_\text{end}}^\text{VMS,e}.\label{eqn:vms_cyc}
		\end{align}
	\end{subequations}
	The analogy between a virtual storage and the flexible operation can be inferred by comparing the summation of (\ref{eqn:vms_bal}) for all time steps $\{0,\dots,T_\text{end}\}$ with the equation (\ref{eqn:vms_cyc}), leading to the following balance:
	\begin{equation}
	\sum_{t\in\mathcal{T}} \mu_{i,t}^\text{VMS,dch} = \sum_{t\in\mathcal{T}}\mu_{i,t}^\text{VMS,ch}.
	\end{equation}
	This balance ensures that, in the flexible charging mode, the optimization problem solver can decide on the optimal charging schedule while preserving the energy balance. 
	As seen in Fig. \ref{fig:bevmodes}, the difference between these lies simply at the defined time step of the demand and the allowance of the VMS. For the flexible charging mode, the demand has to be satisfied at the end of the overnight charging schedule (e.g., 7:00--8:00):
	\begin{align}
	d_{i,t}^\text{mob} &=\begin{cases}
	d_{i}^\text{mob,daily} &\forall t  \in \{\text{7:00--8:00}\}\\
	0 &\forall t \in \{\mathcal{T}\setminus \text{7:00--8:00}\}\end{cases}\label{eqn:mob_dem3}
	\end{align}
	whereas in the inflexible charging mode, the VMS is disabled, and the demand has to be covered fully and precisely at the time of arrival (e.g., 18:00--19:00):
	\begin{subequations}
		\begin{align}
		d_{i,t}^\text{mob} &=\begin{cases}
		d_{i}^\text{mob,daily} &\forall i \in \{\text{18:00--19:00}\}\\
		0 &\forall i \in \{\mathcal{T}\setminus \text{18:00--19:00}\}\end{cases}\label{eqn:mob_dem1}\\
		\mu_{i,t}^\text{VMS,ch} & = \mu_{i,t}^\text{VMS,dch} = 0.\label{eqn:mob_dem2}
		\end{align}
	\end{subequations}

        This concludes the modeling of the building components, their associated flexibilities, and the interactions between them. These buildings are then connected with each other by the LV distribution grid, using the model formulation presented in the following subsection.   
	\subsection{Modeling of the LV distribution grid}\label{sec: grid_comp}
\begin{figure*}[!t]
		\centering
		\includegraphics[width=0.78\linewidth]{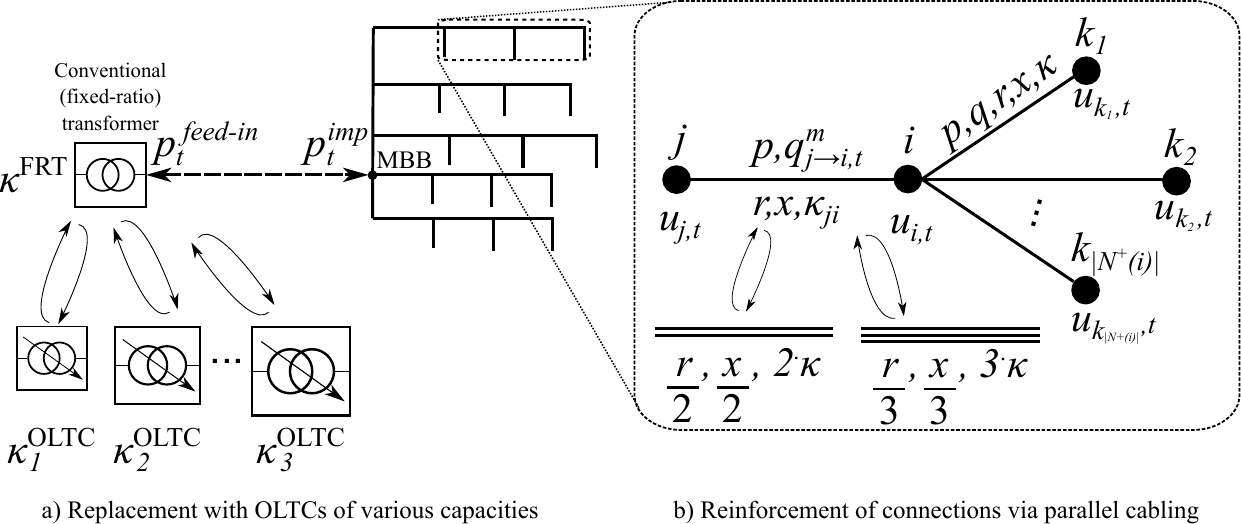}
		\caption{Distribution grid reinforcement measures as defined in the model.}
		\label{fig:grid_exp}
	\end{figure*}

	As the greater focus in this model lies on the analysis of the grid reinforcement measures, the expansion of the grid components is modeled in relatively greater detail compared to that of the building components. This section begins with a description of the grid reinforcement modeling under operational constraints.
	Figure \ref{fig:grid_exp} illustrates the two types of grid reinforcement measures that are allowed in this model: i) addition of parallel lines and ii) replacing the conventional, fixed-ratio transformer (FRT) with an on-load tap changing transformer (OLTC), possibly having a larger capacity. These measures have been assigned high practical relevance for alleviating grids in the rise of PV uptake, as shown in a survey made with ten German grid operators \cite{BenjaminBayer.2018}. The other measures mentioned in the survey, such as wide-area control, involve MV-LV grid interactions, which are not covered by the exclusively LV-grid scope of this work. Similarly, local segmentation of the grid was not included as a tractable formulation was not achievable within the proposed MILP formulation.
	
	\subsubsection{Replacement of the FRT with an OLTC} 
	Replacing the FRT with a larger OLTC not only allows higher capacities for power transfer into and from the grid, but also facilitates voltage regulation through variable tap ratios. OLTCs' utilization for voltage regulation requires real-time nodal voltage measurements \cite{OytunBabacan.2017}, which are assumed to be either present in the grid or are introduced with the OLTC investment. The added degree of freedom through the OLTC increases the allowable range for over- and under-voltages. In order to model the decision of replacing an existing FRT that has a capacity of $\kappa^\text{FLT}$, with a given set of OLTC options $\mathcal O$ having capacities of $\kappa^{\text{OLTC}}_o$ for each OLTC $o\in\mathcal O$, corresponding binary variables $\alpha^\text{FRT}$ and $ \alpha^{\text{OLTC}}_o\; \forall o \in \mathcal O$ are defined first. To ensure the mutual exclusivity of using only one transformer in the planning, we introduce:
	\begin{equation}
	\alpha^\text{FRT}+ \sum_{o\in \mathcal{O}} \alpha^{\text{OLTC}}_o = 1. \label{eqn:one_trafo}
	\end{equation}
	Note that parallelly connected OLTCs can still be modeled by defining a transformer type $o$ that represents a bundle of OLTCs with the respective collective capacity of $\kappa^{\text{OLTC}}_o$. However, this formulation restricts the model to having only a single point of connection between the LV and the MV grid. 
	
	The capacity of the built transformer to inject (and absorb) active power into ($p^{\text{imp}}_t$) and from ($p^{\text{feed-in}}_t$) the distribution grid over the main busbar (MBB) of the grid is then tied to the investment decision as follows:
	\begin{subequations}
		\begin{align}
		p^\text{imp}_t \leq \alpha^{\text{FRT}}\cdot \kappa^\text{FLT}+ \sum_{o\in \mathcal{O}} \alpha^{\text{OLTC}}_o\cdot \kappa^{\text{OLTC}}_o,\label{eqn:ontcap1}\\
		p^\text{feed-in}_t \leq \alpha^{\text{FRT}}\cdot \kappa^\text{FLT}+ \sum_{o\in \mathcal{O}} \alpha^{\text{OLTC}}_o\cdot \kappa^{\text{OLTC}}_o.\label{eqn:ontcap2}
		\end{align}
	\end{subequations}
	As mentioned, OLTCs also assist the grid in respecting the over- and under-voltage limits. For instance, in Germany, the DIN EN 50160 norm requires that the grid voltages are kept within the permissible range of $\pm 10\%$ deviations around the rated voltage $V_\text{base}$ of 400V, at least in 95\% of all 10-minute operation intervals, on a weekly basis\footnote{For simplicity, we enforce this limit to each model time step separately. Yet, this is a conservative approach, and a linear formulation representing the precise regulation is also possible.}. However, this range is reserved for the whole medium and low-voltage ensemble. As the voltages between both grid levels are coupled to each other across FRTs, a more conservative band ($[V'_\text{min},V'_\text{max}]$) is conventionally allocated to the LV grid alone. Based on various norms such as VDE AR-N 4105 and DIN-EN 50160, and considering safety reserves, a voltage band between $+3\%$ and $-5\%$ p.u. may be allocated to the LV grid level \cite{GeorgKerber.2009}. Here, OLTCs grant an additional degree of freedom by adjusting the tap ratio across the transformer (and hence the voltage levels at the MV interface of the LV grid). This decouples the MV and LV grid voltages, allowing them to be set independently from each other. This way, the LV grid can utilize the entire permissible voltage range of $[V_\text{min},V_\text{max}]$. In reality, the discrete nature of the tap change steps or specific requirements from the grid operator may result in further restrictions over the exploitable voltage range \cite{MaximilianArnold.2019}. This behavior is not represented in this model due to its necessity for additional integer variables. 
	
	Table \ref{tab:vband} illustrates the resulting voltage bands for each case and bus. These limitations are imposed on the corresponding bus voltages $u_{i,t} = (V_{i,t})^2$ via the following constraints:
	\begin{subequations}
		\begin{align}
		\dfrac{u_{i,t} }{(V_\text{base})^2}&\leq \left((V'_\text{max})_i^2 +\sum_{o\in \mathcal{O}} \alpha^{\text{OLTC}}_o\cdot \left((V_\text{max})_i^2 - (V'_\text{max})_i^2\right)\right),\label{eqn:vol1}\\
		\dfrac{u_{i,t} }{(V_\text{base})^2}&\geq \left((V'_\text{min})_i^2 +\sum_{o\in \mathcal{O}} \alpha^{\text{OLTC}}_o\cdot \left((V_\text{min})_i^2 - (V'_\text{min})_i^2\right)\right).\label{eqn:vol2}
		\end{align}\label{eq:voltagelimits}
	\end{subequations}
	\begin{table}[h]\footnotesize
		\centering
		\caption{Example permissible voltage bands in the LV grid (per $V_\text{base}$).\label{tab:vband}}
		\setlength{\tabcolsep}{1mm}
		\begin{tabular}{ccccccccc}
			&  Buses $\rightarrow$ & MBB & 1 & 2 & $\dots$ & $n$ & $\dots$ & $\mathcal{N}$\\  \hline
			\multirow{2}{*}{\parbox[c]{1.8cm}{Without OLTC}} & \multicolumn{1}{c|}{$(V'_\text{min})_i$} & \multicolumn{1}{c|}{1} & \multicolumn{6}{c}{0.95} \\ 
			& \multicolumn{1}{c|}{$(V'_\text{max})_i$} & \multicolumn{1}{c|}{1} & \multicolumn{6}{c}{1.03} \\ 
			\multirow{2}{*}{\parbox[c]{1.8cm}{With OLTC}} & \multicolumn{1}{c|}{$(V_\text{min})_i$} & \multicolumn{1}{c|}{0.9} & \multicolumn{6}{c}{0.9} \\
			&\multicolumn{1}{c|}{$(V_\text{max})_i$} & \multicolumn{1}{c|}{1.1}& \multicolumn{6}{c}{1.1} 
		\end{tabular}
	\end{table}
	\subsubsection{Parallel cable reinforcement}
	The grid-relieving functions of parallel lines are twofold: additional parallel lines not only increase the power transfer capacity by increasing the effective conductor area, but also the voltage drop for a given loading is reduced by a smaller effective impedance. The reinforcement of the connection $ji\in \mathcal{L}$ between a given pair of buses $j,i$ is realized by installing a second or third parallel cable. The single, double, and triple cable settings for a given cable type are denoted by $\{$I, II, III$\}$. Here a formulation akin to \cite{JulioLopez.2020} is employed and, similar to transformers, these cable settings are modeled as discrete, mutually exclusive options $\alpha_{ji}^\text{I},\alpha_{ji}^\text{II},\alpha_{ji}^\text{III}$ for each line section $ji$: 
	\begin{equation}
	\sum_{\mathclap{m\in\{\text{I,II,III}\}}} \alpha_{ji}^m = 1.\label{eqn:onecable}
	\end{equation}
    The effective power flow between two buses is then given as the sum over all possible cable settings:
    \begin{subequations}
    \begin{align}
    p_{j\rightarrow i,t}^\text{grid}=\sum_{\mathclap{m\in\{\text{I,II,III}\}}} p_{j\rightarrow i,t}^{m},\label{eqn:totalflowovercables1} \\
    q_{j\rightarrow i,t}^\text{grid}=\sum_{\mathclap{m\in\{\text{I,II,III}\}}} q_{j\rightarrow i,t}^{m}.\label{eqn:totalflowovercables2}     
    \end{align}
    \end{subequations}
	Assuming a single cable of this type has i) a thermal apparent power capacity of $\kappa^\text{cable}$ derived from the maximum allowable current, ii) a reactance of $x^\text{cable}$ and iii) a resistance of $r^\text{cable}$, the corresponding electrical parameters of each setting can be calculated as follows:
	\begin{subequations}
		\begin{align}
		\kappa_{ji}^\text{I} &= \kappa^\text{cable},&& r_{ji}^\text{I} = r^\text{cable},&&x_{ji}^\text{I} = x^\text{cable},\\
		\kappa_{ji}^\text{II} &= 2 \cdot \kappa^\text{cable} ,&&r_{ji}^\text{II} = \dfrac{r^\text{cable}}{2},&& x_{ji}^\text{II} = \dfrac{x^\text{cable}}{2},\\ 
		\kappa_{ji}^\text{III}& = 3 \cdot \kappa^\text{cable},&&r_{ji}^\text{III} = \dfrac{r^\text{cable}}{3},  &&x_{ji}^\text{III} = \dfrac{x^\text{cable}}{3}.
		\end{align}	
	\end{subequations}
 
      The thermal apparent power capacities $\kappa^\text{cable}$ are calculated using the thermal current limits according to the German DIN VDE 0276-60 norm. This calculation also includes a maximum loading limit of 70\%, in conformity with the DIN VDE 0298-4 norm for cables with three loaded conductors. This accounts for a certain level of safety margin in cable dimensioning, as is common practice in grid planning. 
     
	Approximations on the non-convex power flow equations are made to formulate the load flows within a convex optimization problem. Assuming a radial and balanced network with low losses, the LinDistFlow model \cite{MesutBaran.1989} is used for calculating voltage drops over each line section $ji$:
	\begin{align}
	u_{i,t}  &= u_{j,t}  -2\cdot\sum_{\mathclap{m\in\{\text{I,II,III}\}}} \left( r_{ji}^{m}\cdot p_{j\rightarrow i,t}^{m} + x_{ji}^{m}\cdot q_{j\rightarrow i,t}^{m}    \right).\label{eqn:lindistflow}
	\end{align}
	Note that although the sum expression enables a generalized constraint over all possible cable settings, only a single setting (and hence power flow through that setting) is allowed at the solution through the equation (\ref{eqn:onecable}).
 
	The line capacities restrict the active and reactive power through the maximum apparent power: 
	\begin{align}
	\left(p_{j\rightarrow i,t}^{m}\right)^2 + b_y \cdot \left(q_{j\rightarrow i,t}^{m}\right)^2 \leq \left(\kappa_{ji}^m\right)^2 \cdot \alpha_{ji}^m \label{eqn:line_lim0}
	\end{align} 
        Mixed-integer problems with convex quadratic constraints such as (\ref{eqn:line_lim0}) belong to a class known as mixed-integer quadratically constrained problems (MIQCP), and they can be solved by commercial solvers. However, they scale badly in the presence of multi-period models with investment planning. Therefore, the linearized formulation proposed in \cite{DanielAlfonsoContrerasSchneider.2021} and \cite{YueWang.2018} is adopted, with the angle intervals of $\frac{\pi}{4}$: 
	\begin{align}
	a_y \cdot p_{j\rightarrow i,t}^{m} + b_y \cdot q_{j\rightarrow i,t}^{m} \leq c_y\cdot\kappa_{ji}^m \cdot \alpha_{ji}^m \label{eqn:line_lim} \\
	\forall m \in \{\text{I,II,III}\},\;\; \forall y \in \{1,2,\dots, 8\}\;\; \;\;  \nonumber
	\end{align}
	where $a_y,b_y,c_y\; \forall y \in \{1,2,\dots, 8\}$ compose the set of coefficients that define the regular octagon inscribed inside the original feasible space (See Appendix \ref{app: pq}).
	
	Besides these grid reinforcement measures, a central reactive power compensation unit is installed in the main busbar (MBB) for voltage support within the grid, with a variable cost of $c^\text{Qcomp}_{\text{spec},t}$ per kVAr compensated. 
 
    This concludes the modeling of the grid reinforcement options, along with the respective operational constraints and the power flow formulation. Together with the building component constraints introduced in Section \ref{sec: bui_comp}, they define the solution space for HOODS, which can be formulated into an optimization problem for minimizing the total costs associated to all system components.

	\subsection{System costs}\label{sec: costs}
	The total costs incurred for the investment and dispatch of the system components are to be minimized by the optimization solver. Each cost component is listed in Table \ref{tab:costs}. The costs that accrue in each building are associated with electricity provision and the investment/maintenance of the building components. The grid-side costs include those associated with the line and transformer reinforcement and the reactive power compensation. Additionally, through feed-in of excess PV electricity, revenues can be collected in the form of a fixed feed-in tariff. 
 
        For PV and heat pump technologies, the annualized investment costs $c_i^\text{PV},c_i^\text{HP}$ are constituted as
	\begin{equation}
	c_{i}^{x} =af^{x}\cdot \left( c_\text{inv,fix}^x\cdot \beta_i^x  + c_\text{inv,var}^{x}\cdot\kappa_i^x \right) + c_\text{O\&M}^{x} \cdot \kappa_i^x \;\; \forall x \in \{\text{PV}, \text{HP}\}\label{eqn:pvhpcosts}
	\end{equation}
	where $af^{x} $ is the annuity factor\footnote{The annuity factor is calculated for each component as $af = \dfrac{i\cdot(i+1)^N}{(i+1)^N-1}$, where $i$ and $N$ stand for the weighted average capital of cost (WACC) for and the economic lifetime of the component respectively.} of the component to annualize the investment costs, $c_\text{inv,fix}$ and $c_\text{inv,var}$ the fixed and capacity-dependent specific investment costs, and $c_\text{O\&M}^{x}$ the yearly operation and maintenance costs. The running costs of the system are limited to the total electricity import into the distribution grid with the price $c^\text{imp}_{\text{spec},t}$ and it is assumed that the system can be renumerated by $c^\text{feed-in}_{\text{spec},t}$ for each unit of exported PV electricity. Although the retail electricity price, in reality, contains a network surcharge that serves to recover the costs from grid investment and thus correlates with it (various surcharge-setting options as investigated in \cite{Schittekatte.2018}), an analysis of such endogenous interaction is outside the scope of this work. Instead, constant volumetric rates are assumed.
 
 For the thermal storage unit, it is assumed that the energy and power capacities can be invested independently (in contrast to the battery storage, where a fixed energy-to-power ratio is set). In order to ensure that the simultaneous charging and discharging of batteries ("unintended storage cycling" \cite{MartinKittel.2022}) do not take place at the same time, a low variable cost $c^{\text{sto},\epsilon}_\text{spec}=0.001$ \euro/kW is attached to the storage operation:
   \begin{align}
   &c^{\text{sto},\epsilon}_{i} = c^{\text{sto},\epsilon}_\text{spec}\cdot \\&\sum_{t\in\mathcal{T}}\left(p_{i,t}^\text{bat,dch}+p_{i,t}^\text{bat,ch}\nonumber\vphantom{x}+\gamma_{i,t}^\text{TS,dch}+\gamma_{i,t}^\text{TS,ch}+\mu_{i,t}^\text{VMS,dch}+\mu_{i,t}^\text{VMS,ch}\vphantom{\mu_{i,t}^\text{VMS,ch}}\right).\label{eqn:epsilonstocosts}
   \end{align}
   While this cost component is too small to have a significant influence on the optimal operation, it prevents the unintended storage cycling without using an integer formulation with high computational complexity.
 
    The costs for cables consist of the installation and material costs, with both depending on the line length, i.e. 
	\begin{subequations}
		\begin{align}
		c_{ji,\text{spec}}^{\text{line},\text{II}} &= l_{ji}\cdot \left(c_\text{ins} + c_\text{mat}\right),\\
		c_{ji,\text{spec}}^{\text{line},\text{III}} &= l_{ji}\cdot \left(c_\text{ins} + 2\cdot c_\text{mat}\right),
		\end{align}
	\end{subequations}
	where $c_\text{ins}, c_\text{mat}$ stand for the aforementioned cost components per length of the cable section $l_{ji}$. Distinct unit costs for each OLTC $o$ are defined as $c^{\text{OLTC}}_o$, which include the cost of the transformer, the secondary units (e.g. controllers) and the installation thereof.
	\renewcommand{\arraystretch}{1.4}		
	\begin{table}[h]\footnotesize
		\centering
		\caption{Cost representations for each system component.}
		\setlength{\tabcolsep}{1mm}
		\begin{tabular}{l|cc}
			Costs &  Buildings $i\in \mathcal{I}^b$ & Grid$^*$ \\ \hline 
			$c_i^\text{PV}$ & (\ref{eqn:pvhpcosts})&  \\
			$c_i^\text{bat}$ & $c_\text{spec}^\text{bat,c}\cdot\kappa_i^\text{bat,c}$&  \\
			$c_i^\text{HP}$ & (\ref{eqn:pvhpcosts})&  \\
			\multirow{2}{*}{\parbox[c]{0.4cm}{$c_i^\text{TS}$}} & $c_\text{spec}^\text{TS,p}\cdot\kappa_i^\text{TS,p}$ &  \\
			& $+c_\text{spec}^\text{TS,c}\cdot\kappa_i^\text{TS,c}$&  \\
            $c^{\text{sto},\epsilon}_i$ & (\ref{eqn:epsilonstocosts}) &\\ 
            $c^\text{imp}$ & $*$  & \multicolumn{1}{l}{$\phantom{-}\sum\limits_{\mathclap {t}\in \mathcal{T}}\left(\omega_t \cdot c^\text{imp}_{\text{spec},t} \cdot  p^{\text{imp}}\right)$} \\
			$c^\text{feed-in}$ & $*$ & \multicolumn{1}{l}{$-\sum\limits_{\mathclap{t}\in \mathcal{T}}\left(\omega_t \cdot c^\text{feed-in}_{\text{spec},t} \cdot  p^{\text{feed-in}}_t\right)$}  \\
			$c^\text{Qcomp}$ & &   \multicolumn{1}{l}{$\phantom{-}\sum\limits_{\mathclap {t}\in \mathcal{T}}\left(\omega_t \cdot c^\text{Qcomp}_{\text{spec},t} \cdot  q^{\text{comp}}_t\right)$}  \\			
			$c^{\text{line}}$ &  &   \multicolumn{1}{l}{$\phantom{-}\sum\limits_{{ji}\in \mathcal{L}}\left(\sum\limits_{m\in \{\text{II},\text{III}\}}\left(c_{ji,\text{spec}}^{\text{line},m}\cdot\alpha_{ji}^{m}\right)\right)$}\\
			$c^{\text{OLTC}}$ &  &   \multicolumn{1}{l}{$\phantom{-}\sum\limits_{\mathclap{{o}\in \mathcal{O}}}\left(c_{o}^{\text{OLTC}}\cdot \alpha^{\text{OLTC}}_o\right)$}\\
			\bottomrule	
			\multicolumn{3}{l}{$^*$\footnotesize In uncoordinated paradigms, these costs are distributed into all }\\			\multicolumn{3}{l}{buildings by their individual import and feed-ins, see \textsc{HOODS-Bui}.}
		\end{tabular}\label{tab:costs}
	\end{table}
	\renewcommand{\arraystretch}{1}

 	\subsection{HOODS model formulation}\label{sec: modfor}
	With the above cost representations and the operational constraints combined, the LV grid/building system expansion and operation planning model (denoted by \textsc{HOODS}) can be formulated as an optimization problem in the following form:
	
	\textsc{HOODS}($\mathcal{T},$ flexible):
	\begin{align*}
	\min_{\mathclap{\substack{\boldsymbol \kappa, \boldsymbol \epsilon, \boldsymbol \alpha, \boldsymbol f}}} \;\;\;\sum_{\mathclap{i\in\mathcal{I}}}&(c_i^\text{PV}+c_i^\text{bat}+c_i^\text{HP}+c_i^\text{TS}+c^{\text{sto},\epsilon}_i)\nonumber\\ &+c^\text{imp}+c^\text{feed-in}+c^\text{Qcomp}+c^{\text{line}} + c^\text{OLTC}\\
	\text{s.t.} & \begin{multlined}[t] \{(\ref{eqn:pv}-\ref{eqn:pq}),(\ref{eqn:bat_bal}-\ref{eqn:bat_cap_e}) ,(\ref{eqn:hpcop}-\ref{eqn:hpcap}),\\(\ref{eqn:ts_bal}-\ref{eqn:ts_cap_e}), (\ref{eqn:heat_balance}), (\ref{eqn:bevconv}-\ref{eqn:bevcap}), (\ref{eqn:mobbal}),(\ref{eqn:vms_bal}-\ref{eqn:vms_cap_ch})  ,\\ (\ref{eqn:vol1}-\ref{eqn:vol2}) \hspace{1.4cm} \forall i \in \mathcal{I}^b, \quad\forall t \in \mathcal{T}\} \nonumber \end{multlined}  \\
 & (\ref{eqn:pbal}-\ref{eqn:qbal}) \hspace{3cm}\forall i \in \mathcal{I}, \quad\forall t \in \mathcal{T}\\
	&  (\ref{eq: pvlimit}),(\ref{eqn:bat_cyc}-\ref{eqn:bat_e2p}),(\ref{eq: hplimit}),(\ref{eqn:ts_cyc}),(\ref{eqn:vms_cyc})\hspace{1.5cm}\forall i \in \mathcal{I}^b\nonumber  \\
	&  (\ref{eqn:onecable}) \hspace{5cm}\forall ji \in \mathcal{L}\nonumber  \\
	&  (\ref{eqn:totalflowovercables1}-\ref{eqn:totalflowovercables2}),(\ref{eqn:lindistflow}) \hspace{1.5cm} \forall ji \in \mathcal{L}, \quad\forall t \in \mathcal{T}\nonumber  \\
	&  (\ref{eqn:ontcap1}-\ref{eqn:ontcap2}) \hspace{4cm} \forall t \in \mathcal{T}\nonumber  \\
	&  (\ref{eqn:line_lim})\quad\forall t \in \mathcal{T},\forall m \in \{\text{I,II,III}\}, \forall y \in \{1,2,\dots,8\}\nonumber \\
	&  (\ref{eqn:one_trafo}) \nonumber \\
	 &\textbf{if} \text{ flexible}: \hspace{0.3cm}(\ref{eqn:mob_dem3}),\hspace{1.4cm} \forall i \in \mathcal{I}^b, \quad\forall t \in \mathcal{T}\nonumber\\
	&\textbf{else}: \{\begin{multlined}[t] \;\;(\ref{eqn:mob_dem1}-\ref{eqn:mob_dem2}),\;\; \; \kappa_i^\text{TS},\kappa_i^\text{bat},\kappa_i^\text{VMS} = 0\\ \forall i \in \mathcal{I}^b, \quad\forall t \in \mathcal{T}\} \nonumber \end{multlined}
	\end{align*}
	where, for brevity, the capacity variables are grouped into 
	\begin{equation}
	\boldsymbol \kappa =\begin{bmatrix}\kappa_i^\text{PV},\beta_i^\text{PV},\kappa_i^\text{bat,e},\kappa_i^\text{bat,p},\kappa_i^\text{HP},\beta_i^\text{HP},\kappa_i^\text{TS,e},\kappa_i^\text{TS,p}&\forall i \in \mathcal{I}^b\end{bmatrix},
	\end{equation}
	the building operation variables into 
	\begin{align}
	\boldsymbol \epsilon =\begin{bmatrix*}[l]
	p_{i,t}^\text{PV},q_{i,t}^\text{PV},p_{i,t}^\text{bat,ch},p_{i,t}^\text{bat,dch},\epsilon_{i,t}^\text{bat,ch},&\\
	p_{i,t}^\text{HP},q_{i,t}^\text{BEV},\gamma_{i,t}^\text{HP},\gamma_{i,t}^\text{TS,dch},&\forall i \in \mathcal I^b, \forall t \in \mathcal{T}\\
	\gamma_{i,t}^\text{TS,ch} \epsilon_{i,t}^\text{TS,e},\mu_{i,t}^\text{BEV},\mu_{i,t}^\text{VMS,dch},\mu_{i,t}^\text{VMS,ch}&
	\end{bmatrix*},
	\end{align} the grid expansion variables into 
	\begin{equation}
	\boldsymbol \alpha = \begin{bmatrix*}[l]
	\alpha_{ji}^m &\forall ji \in \mathcal L, \forall m \in \{\text{I,II,III}\}\\
	\alpha^\text{FRT}&\\
	\alpha^\text{OLTC}_o\quad&\forall o \in \mathcal O
	\end{bmatrix*}
	\end{equation} and the grid operation variables into 
	\begin{align}
	\boldsymbol f = \begin{bmatrix*}[l]
	u_{i,t}&\forall i \in \mathcal I, \forall t \in \mathcal T\\  p_{{j\rightarrow i},t}^m,q_{{j\rightarrow i},t}^m&\forall ji \in \mathcal{L}, \forall m \in \{\text{I,II,III}\},  \forall t \in \mathcal T \\
	p_t^\text{imp}, p_t^\text{feed-in}, q_t^\text{comp} &\forall t \in \mathcal T
	\end{bmatrix*}.
	\end{align}
  	Note the model setting "flexible", which allows investing in the storage components and intelligent charging of the BEVs.
   
	Table \ref{tab: inputoutput} summarizes the inputs and outputs of the HOODS model. The corresponding optimization problem has a mixed-integer linear formulation and can be solved by any off-the-shelf solver. Thereby, a holistic optimization for the whole DS can be performed, which corresponds to a compromise solution that caters best to maximizing the overall welfare of all agents in the DS. In this process, it assumes full cooperation between the system participants and hence no conflicts of interest between them. Additional to this, further assumptions were made to constitute the problem into the presented MILP model formulation. The major model assumptions are summarized below:
 \begin{itemize}
 \item the single-year modeling approach, while grid planning usually takes multi-year periods in account, 
\item a balanced distribution grid with a radial structure, 
\item perfect forecast of the HEMS operation for the flexibility supply. Most HEMS work with either rule-based or at most model predictive control algorithms with limited and imperfect forecast in practice. Nevertheless, discrepancies are limited due to short-term (daily) operation of the storages, as they are mainly utilized to integrate the diurnal fluctuations of the PV electricity.
 \item no consideration of stochasticity in model parameters such as demands, component costs and PV availability. Reliability of the planning is instead achieved via safety margins in cable loading and voltage limitations,
 \item the limitation of grid reinforcement measures to those elaborated, 
 \item a constant, volumetric retail price for electricity, whereas dynamic tariffs are becoming increasingly common in Europe \cite{EuropeanCommission.2019}, and 
 \item the linear modeling of the model components as described.
 \end{itemize}

\begin{table}
\centering \caption{Inputs and outputs of the HOODS model.}
\resizebox{\columnwidth}{!}{%
\begin{tabular}{p{0.5\columnwidth}|p{0.5\columnwidth}}
Model Inputs                                         & Model Outputs                           \\ \hline
1) Grid data: topology and parameters                       & 1) Costs: investment \& operation \\
2) Building demand data: electricity, heat and mobility   & 2) Capacity and operation planning of building components                                               \\
3) Techno-economic parameters for the building and grid components & 3) Nodal voltages and load flow conforming to grid restrictions                                                          \\
4)  PV capacity factor \& HP COP profiles                                       & 4) Grid reinforcement measures: OLTC and cables\\ 
5) Electricity prices: active and reactive components & 5) Grid-side curtailment schedule            \\                                                 &                                        
\end{tabular}\label{tab: inputoutput}
}\end{table}

        \section{Planning paradigm definitions based on HOODS} \label{sec: paradigms}
        The HOODS model formulation presented in the previous section allows a social planner to flexibly optimize the costs for the entire DS. However, issues regarding the tractability of the problem and the practicality of its assumptions under the current regulations arise. 
        The computational complexity of the HOODS problem depends on the temporal resolution and the time horizon of the model. This issue can be alleviated by the aggregation of model time series. Moreover, an analysis of several variations on the HOODS model is required, which should reflect different levels of flexibility utilization and coordination between the DS agents. The following subsections describe these two ideas in detail.
        
	\subsection{Bi-level holistic modeling with varying temporal aggregation for complexity reduction (\textit{Coor$_+$Flex$_+$} and \textit{Coor$_+$Flex$_-$})}
	The MILP formulation of the grid reinforcement measures enables the representation of realistic discrete decisions (i.e., replacing the transformer) and variable impedance effects (through cable reinforcement). However, it does not scale well numerically, even in commercial solvers such as Gurobi or CPLEX, for many modeled time steps and buildings. Hence, a bi-level approach is applied to reduce the computational complexity of the HOODS problem. First, the capacity planning is done by considering a reduced time series $\mathcal{T}_r$ that consists of four typical weeks with corresponding weights (occurrences) for each time step $\omega_t$, instead of modeling every hour of the year. A hierarchical clustering of the time series within the model is made using the \textit{tsam} module \cite{MaximilianHoffmann.2022}. 
 
 This sequential approach for dimensioning and operation, while necessary due to computational limitations, indeed leads to a deviation from the globally optimal system capacities. However, the difference is expected to be limited, as the resulting aggregation led to an acceptable accuracy in representing the primary demand curves for the case study introduced in Section \ref{sec: casestudy}. As shown in Figure \ref{fig:tsam_accuracy} in the appendix, the peak and mean values of the aggregated electricity and space heating demand duration curves coincide with the original time series. As these two metrics are significant for capacity planning, the aggregated model can serve as a suitable proxy for the optimal dimensioning of the components. After the capacities are optimized in the first stage, these are fixed in a consequent non-aggregated operational optimization model to generate hourly operation curves for the whole year (See Figure \ref{fig:paradigms}d, and Algorithm \ref{alg1} in Appendix \ref{app: alg}). By following this approach, the tractability of the HOODS problem is maintained for representative scales of LV distribution grids. This bi-level coordinated planning paradigm is henceforth denoted by \textit{Coor$_+$Flex$_+$}. To investigate the individual benefit of system-wide coordination and local energy exchange, a variation of this paradigm that excludes all flexibility options is also made, which is denoted as \textit{Coor$_+$Flex$_-$} (See Figure \ref{fig:paradigms}b, and Algorithm \ref{alg4} in Appendix \ref{app: alg}).
	
	\subsection{Multi-level sequential modeling with varying levels of flexibility (\textit{Coor$_-$Flex$_-$} and \textit{Coor$_-$Flex$_+$})}\label{sec: uncoord_paradigms}
	Besides the computational complexity mentioned above, a further issue with the HOODS formulation lies in its supposition of a single entity that has i) access to all of the information at the distribution grid level (e.g., the exact building stock, the energy consumption profiles) and ii) authority over the integrated planning of all system components, the ensuing operation thereof, and the local energy exchange between households. This supposition is far from the current reality. The adoption of DERs at a building level is highly motivated by individual interests rather than system benefits. Moreover, the exchange of PV electricity between buildings in times of excess to increase its overall integration, as this framework allows, requires advanced local markets, which are not an established practice yet (despite examples of peer-to-peer (P2P) markets as described in \cite{PioBaake.2020}). Barring pilot projects, the majority of distribution grids, even in developed countries such as Germany, lack the communication protocols, measurement instruments, and regulatory framework necessary for an intelligent operation that would approximate the social optimum \cite{WolfgangZander.2018}. 
	
\begin{figure*}[t]
	\centering
	\includegraphics[width=\linewidth]{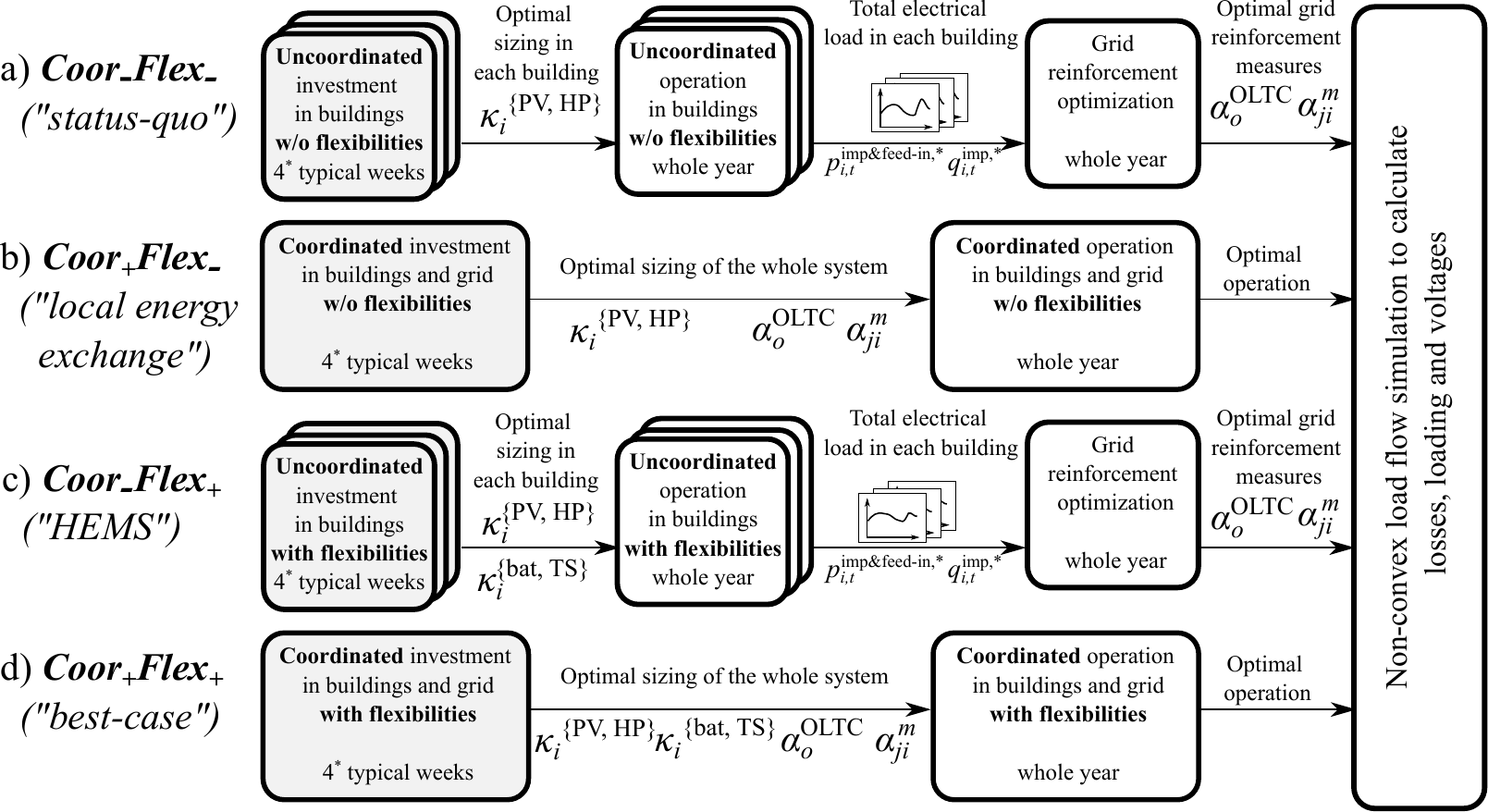}
	\caption{The investigated planning paradigms on the distribution grid level.}
	\label{fig:paradigms}
\end{figure*}

	Therefore, the results of the coordinated \textit{Coor$_+$Flex$_+$} paradigm are rather considered the benchmark for the best-case planning and operation of the whole system. To be compared with this benchmark, two further variations on this paradigm are implemented (see Figures \ref{fig:paradigms}a and \ref{fig:paradigms}c). On the one hand, these variations approximate the status-quo practice, dropping the assumption of holistic system coordination and replacing it with a sequential process. On the other hand, they allow a highly scalable formulation as the models solving the building (sub)problems are now independent. Note that it is not aimed to simulate a specific market mechanism through these sequential approaches. Rather, they are used to quantify the welfare losses that are suffered in practical system operation, where flexibilities may not be exploited to their fullest potential due to a lack of foresight and coordination. 
 
    The procedure for the uncoordinated paradigms is as follows: first, the individual building components are optimally dimensioned to minimize their costs (denoted by the \textsc{HOODS-Bui} subproblems). As these building-specific problems are grid-unaware, the variables for the network flows interfacing each building are substituted with building-specific import and feed-in variables $\boldsymbol{p}_{i}^\text{import}, \boldsymbol{p}_{i}^\text{feed-in}, \boldsymbol{q}_{i}^\text{import}$ in the power balance constraints (\ref{eqn:pbal}--\ref{eqn:qbal}):
	\begin{align}
	&\begin{multlined}
	p_{i,t}^\text{PV} + p_{i,t}^\text{bat,dch} + p_{i,t}^\text{imp} \\= d_{i,t}^\text{elec} + p_{i,t}^\text{bat,ch} + p_{i,t}^\text{BEV}+ p_{i,t}^\text{HP} + p_{i,t}^\text{feed-in}
	\end{multlined}\tag{\ref{eqn:pbal}'}\\
	&q_{i,t}^\text{PV} + q_{i,t}^\text{imp} = d_{i,t}^\text{elec,q} \tag{\ref{eqn:qbal}'}
	\end{align}
	The resulting component dimensioning problem for a given building $i\in \mathcal I^b$ thus has the following form:
	
	\textsc{HOODS-Bui($\mathcal{T},i,$} flexible):
	\begin{align}
	&\min_{\mathclap{\substack{\boldsymbol \kappa_i, \boldsymbol \epsilon_i, \boldsymbol{p}_{i}^\text{imp},  \boldsymbol{p}_{i}^\text{feed-in}, \boldsymbol{q}_{i}^\text{imp}}}} \;\;\;c_i^\text{PV}+c_i^\text{bat}+c_i^\text{HP}+c_i^\text{TS}+c^{\text{sto},\epsilon}_i+c^\text{imp}_i+c^\text{feed-in}_i\nonumber\\
	&\text{s.t.} \{\begin{multlined}[t] (\ref{eqn:pv}-\ref{eqn:pq}),(\ref{eqn:bat_bal}-\ref{eqn:bat_cap_e}),(\ref{eqn:pbal}'-\ref{eqn:qbal}'), (\ref{eqn:hpcop}-\ref{eqn:hpcap}),\\(\ref{eqn:ts_bal}-\ref{eqn:ts_cap_e}), (\ref{eqn:heat_balance}), (\ref{eqn:bevconv}-\ref{eqn:bevcap}), (\ref{eqn:mobbal}),(\ref{eqn:vms_bal}-\ref{eqn:vms_cap_ch}),\\ (\ref{eqn:vol1}-\ref{eqn:vol2}) \quad\forall t \in \mathcal{T} \}\nonumber \end{multlined} \\
	& \hspace{1cm} (\ref{eq: pvlimit}),(\ref{eqn:bat_cyc}-\ref{eqn:bat_e2p}),(\ref{eq: hplimit}),(\ref{eqn:ts_cyc}),(\ref{eqn:vms_cyc}),\nonumber \\
	&\textbf{if} \text{ flexible}: \hspace{0.3cm}(\ref{eqn:mob_dem3}),\nonumber\\
	&\textbf{else}: \;\;(\ref{eqn:mob_dem1}-\ref{eqn:mob_dem2}),\;\; \; \kappa_i^\text{TS},\kappa_i^\text{bat},\kappa_i^\text{VMS} = 0\nonumber
	\end{align}

	After solving for the cost-optimal dimensioning and operation of the building components, the resultant electrical load and feed-in profiles ($\boldsymbol{p}^\text{imp},  \boldsymbol{p}^\text{feed-in}, \boldsymbol{q}^\text{imp}\;  \forall i \in \mathcal{I}^b$) are communicated to the grid planner. The grid planner can then invest in the necessary reinforcement measures which correspond to the minimal costs while respecting the grid constraints (denoted by the \textsc{HOODS-Grid} subproblem):
	
	{\textsc{HOODS-Grid}}($\mathcal{T},\boldsymbol{p}^{\text{imp},*},  \boldsymbol{p}^{\text{feed-in},*}, \boldsymbol{q}^{\text{imp},*}$):
	\hspace{0.2cm}\begin{align}
	&\min_{\mathclap{\substack{\boldsymbol \alpha, \boldsymbol f}}} \;\;\; c^\text{line} + c^\text{OLTC} + c^\text{Qcomp} + \sum\limits_{\mathclap{{i}\in \mathcal{I}}}\sum\limits_{\mathclap{{t}\in \mathcal{T}}}\left(c^\text{feed-in}_{\text{spec},t}\cdot {p}_{i,t}^\text{PVcurt}\right)  \nonumber\\
&\text{s.t.}  	 (\ref{eqn:onecable}) \;  \;  \;  \; \quad\quad\quad\quad\quad\quad\quad\quad\quad\quad\quad\;\quad\forall ji \in \mathcal{L}\nonumber  \\
&  (\ref{eqn:totalflowovercables1}-\ref{eqn:totalflowovercables2}),(\ref{eqn:lindistflow}) \quad\quad\quad\ \quad\quad\quad \quad\quad\ \forall ji \in \mathcal{L}, \quad\forall t \in \mathcal{T}\nonumber  \\
&  (\ref{eqn:ontcap1}-\ref{eqn:ontcap2}) \quad\quad\quad\quad\quad\quad \quad\quad\quad\quad\;\;\; \forall t \in \mathcal{T}\nonumber  \\
&  (\ref{eqn:line_lim})\quad\forall m \in \{\text{I,II,III}\}, \forall t \in \mathcal{T}, \forall y \in \{1,2,\dots,8\}\nonumber \\
&  (\ref{eqn:one_trafo}) \nonumber \\
&\begin{multlined}[t]\sum_{\mathclap{j \in N^{-}(i) }} p_{j\rightarrow i,t}^\text{grid} - \sum_{\mathclap{j \in N^{+}(i) }}p_{i\rightarrow k,t}^\text{grid} - {p}_{i,t}^\text{PVcurt}= {p}_{i,t}^{\text{imp},*} - {p}_{i,t}^{\text{feed-in},*}  \\   \forall i \in \mathcal{I}, \forall t \in \mathcal T \label{eqn:adj_p}\end{multlined} \\
&\sum_{\mathclap{j \in N^{-}(i) }} q_{j\rightarrow i,t}^\text{grid} - \sum_{\mathclap{j \in N^{+}(i) }}p_{i\rightarrow k,t}^\text{grid} = {q}_{i,t}^{\text{imp},*}\; \;  \;  \;  \;  \quad\forall i \in \mathcal{I}, \forall t \in \mathcal T \label{eqn:adj_q}
\end{align}
The equations (\ref{eqn:adj_p}--\ref{eqn:adj_q}) are the modified nodal balance equations that relate the power flows to the pre-optimized active and reactive power injections for each network bus. Additional to the passive reinforcement measures, the grid operator can remotely curtail an amount of excess PV feed-in that buildings intend to feed into the grid to avoid grid reinforcement. In return, they have to compensate the affected households by the usual feed-in tariff $c^\text{feed-in}_{\text{spec},t}$ per curtailed amount of electricity.

Note that in order to maintain comparability with the coordinated paradigm, the time-series aggregation is kept also in the uncoordinated paradigms for the \textsc{HOODS-Bui} subproblem. This leads to the three-step approach for these paradigms, as depicted in Figure \ref{fig:paradigms}.

Using these subproblem definitions, the uncoordinated planning paradigms are defined as follows.
\subsubsection{Uncoordinated paradigm without flexibilities}\label{sec:uncoor_inflex}
The first uncoordinated paradigm variation, denoted by \textit{Coor$_-$Flex$_-$}, allows no flexibility options within the buildings. Hence, similar to \textit{Coor$_+$Flex$_-$}, the building investment model deals with only the dimensioning of the heat pumps and PV for individual buildings to cover a "rigid" demand.  

\subsubsection{Uncoordinated paradigm with flexibilities}
The second paradigm variation, \textit{Coor$_-$Flex$_+$}, assumes that for each building, flexibility measures can be dimensioned parallel to the generation components and can be used to minimize the building energy costs (e.g., by a home energy management system--HEMS). As the retail electricity prices are exceedingly above the feed-in tariffs in many countries, the building-level benefits of these flexibilities usually manifest themselves as higher self-consumption of PV electricity. 

For the exact formulations of these uncoordinated paradigms, the reader is referred to the Algorithms \ref{alg2} and \ref{alg3} in Appendix \ref{app: alg}, respectively. 
\subsection{Subsequent non-convex power flow simulation}
To maintain the tractability of the optimization problem, a simplified power flow formulation \textit{LinDistFlow} had been used for the grid planning step. This formulation, however, drops the terms related to losses in the power flow and voltage drop equations, which might lead to approximation errors. Therefore, after the system planning is finalized in each paradigm, a non-convex load flow simulation is conducted with the reinforced grid components and the resultant nodal injections. The subsequent power flow simulation serves two concrete purposes:
\begin{itemize}
\item \textit{Calculating the total power losses in the system:} In order to keep the supply/demand balance in the grid, compensation for the transport losses of electricity is procured and paid by the grid operator in Germany \cite{AdolfSchwab.2009}. Correspondingly, these cost terms will be accounted to the grid operator in the economic analysis (Subsection \ref{sec: costs}).
\item \textit{Re-calculating the resultant bus voltages and line loadings:} By recalculating the voltages and the loadings, we ensure the satisfaction of the corresponding restrictions.
\end{itemize}
The power flow calculation is based on Newton-Raphson method and is conducted using the \textit{pandapower}\cite{pandapower.2018} library.
\section{Case study: a village-type distribution grid in Germany} \label{sec: casestudy}
In this section, a demonstration of the HOODS model and the introduced paradigms is made using a case study. The case study deals with the total electrification of all energy demands within a village-type, residential distribution grid in southern Germany. The following sections will describe the scenario data and discuss the results. 
\subsection{Data}
As shown in Table \ref{tab: inputoutput}, HOODS requires a set of input data to parametrize a specific scenario. These consist of grid data, building data, and techno-economic data. 
\subsubsection{Grid data}
Due to the lack of actual distribution grid data, the corresponding representative Kerber distribution grid model is assumed \cite{GeorgKerber.2011}. The key figures of this grid are as follows:
\begin{itemize}
	\item{six branches}
	\item{116 buses, of which 57 are load buses} 
	\item{114 line sections} 
	\item{a fixed-ratio transformer with a capacity of 400 kVA}
	\item{NAYY-J 4x150mm$^2$ cable type for the primary grid}	
	\item{NYY-J cable type for the service lines connecting the building premises to the distribution grid. Service line cross sections are dimensioned depending on the number of dwellings in each building according to the DIN 18015-1 norm, assuming no electrical water heating in the building before switching to heat pumps}	
\end{itemize}
\subsubsection{Building and demand data}
To each load bus of the grid, residential buildings with various properties are assigned. The properties that are influential to energy demand were derived from a synthetic building database that is applied in the scope of Bavaria, Germany. This database combines the following open data sources: 
\begin{itemize}
	\item OpenStreetMaps data\footnote{Accessible under \url{http://download.geofabrik.de/europe/germany/bayern.html}} for building geometry, land use, and points of interest
	\item CORINE Land Cover (CLC) inventory\footnote{Methodology can be found in \cite{GyorgyButtner.2021}, dataset in \url{https://land.copernicus.eu/pan-european/corine-land-cover/clc2018}.} for land use
	\item Census Database 2011 for the number of occupants,
	\item TABULA building typologies for the building typologies in terms of refurbishment, building type, and construction year
\end{itemize}
to obtain synthesized characteristics for any given residential building (such as building type, construction year, plot and living space, number of floors, refurbishment level, number of dwellings, and occupants). A subset of buildings was chosen from a district in Unterhaching, a semi-rural municipality near Munich. Selected buildings span a wide range between one and nine dwellings each and are characterized by around 50 m$^2$ of living space per occupant. The hourly space heating and daily domestic hot water demands\footnote{Daily domestic hot water demands are then split into a subset of hours for each day, depending on the number of dwellings for every building (e.g. a daily water demand of 20 kWh/day for a building with two dwellings is split into two distinct hours with 10 kWh/h each.} are calculated using the open-source UrbanHeatPro software\footnote{Accessible under https://github.com/tum-ens/UrbanHeatPro}, which combines RC- and activity-based models to generate hourly demand profiles for a given building \cite{AnahiMolarCruz.2022}. For the hourly profiles of the active power consumption of home appliances, the activity-based REM workflow \cite{AkhilaJambagi.2021} has been used.

As the overall BEV penetration is currently limited, a long-term electrification scenario is considered to examine the extreme effects of mobility electrification on the grid. Based on a prognosis of 36 million BEVs by 2045 \cite{Prognos.2021}, combined with a corresponding mobility demand of 424 billion person km by 2050 \cite{Prognos.2021}, and a BEV power consumption of around 16 kWh/100 km \cite{OkoInstitut.2014}, a daily BEV consumption of 5.15 kWh/BEV is derived. Aggregate daily mobility demands for each building are then calculated by assigning a BEV for each second occupant. With this approach, however, a simplistic assumption is made that the consumption does not change seasonally or depending on weekdays and weekends. The charging station capacity is pre-defined as 11 kW per BEV, in line with the federal funding for installing controllable stations \cite{GermanFederalMinistryfordigitalandtransport.2022}. In the inflexible paradigm, these daily demands are distributed uniformly between the various arrival times of 17:00, 18:00, and 19:00. Correspondingly, in the flexible paradigms, the charging of VMS is allowed until 7:00, 8:00, or 9:00 the next day depending on the arrival time. 

The resultant demand distributions can be seen in Figure \ref{fig:demand_visual} in the appendix schematically. The average annual demands for electricity and heating are around 2.3 and 11.6 MWh per dwelling, while the electricity demand for mobility is around 1.5 MWh. With these demands, roughly a tripling in the bulk electricity demand due to 100\% electrification is estimated\footnote{Assuming a yearly average COP of 3.5 for the heat pumps (thus ignoring lower COPs for winter days where the heating demand is significantly higher): $(11.6/3.5 + 1.5)/2.3 = 210\% $ increase.}.
\subsubsection{Techno-economic data}
Further model data include the techno-economic parameters concerning the system components, whose values and sources can be found in Table \ref{tab: techeco} in the appendix. 
\subsection{Results of the case study}
This section illustrates and discusses the model results of the case study. First, the system operation in each paradigm is described, leading to grid reinforcement wherever necessary. The section is then concluded with a cost comparison between the paradigms to quantify the benefits of the flexibilities and inter-agent coordination to the social welfare of the system.
\subsubsection{Distribution system operation}
 The first analysis focuses on the hourly distribution system operation to analyze the system behavior in high temporal detail and, consequently, its influence on the grid reinforcement. For a broader picture of the monthly electricity balances throughout the year, the reader is referred to Figure \ref{fig:monthly_results} in the Appendix. Figure \ref{fig:summerwinter} shows the aggregated DS operation, each row representing a five-day period in the summer and winter seasons, respectively.

The winter season is characterized by limited electricity generation from PV and high heating demands. When these conditions are combined, the system becomes dependent to a large extent on imports from the upstream grid. The individual benefit of energy exchange (\textit{Coor$_+$Flex$_{-}$}) is limited in winter, as excess PV production occurs simultaneously in every building. Building-level flexibilities, on the other hand, allow a better integration of the PV electricity, reducing the feed-in considerably. The combination of coordination and flexibilities reveals a synergy effect: the exchange of excess power and, if  necessary, its storage in locally available batteries allow a highly flexible utilization, both spatially and temporally. As a result, even better integration of this electricity is achieved, such that almost no feed-in occurs. Moreover, through the central management of BEV charging, the peak imports are reduced. 

The summer season depicts an almost opposite picture. Here, due to high generation from PV, grid import is only necessary for covering the electrical demand in the evening hours and the BEV charging in the late afternoon (which is only partially incident to the hours with PV electricity). Instead, vast amounts of excess PV electricity are fed into the grid. Through coordinated investment, the system demand can be covered optimally with smaller capacities of PV installations (see lower PV generation for \textit{Coor$_+$Flex$_{-}$} in summer). As flexibilities provide further means of integration in \textit{Coor$_-$Flex$_{+}$} and \textit{Coor$_+$Flex$_{+}$}, the system can tolerate higher capacities of PV again. Thereby, the system reaches a state where the imports are significantly reduced. 
\begin{figure*}[!htb]
	\centering
	\includegraphics[width=1\linewidth]{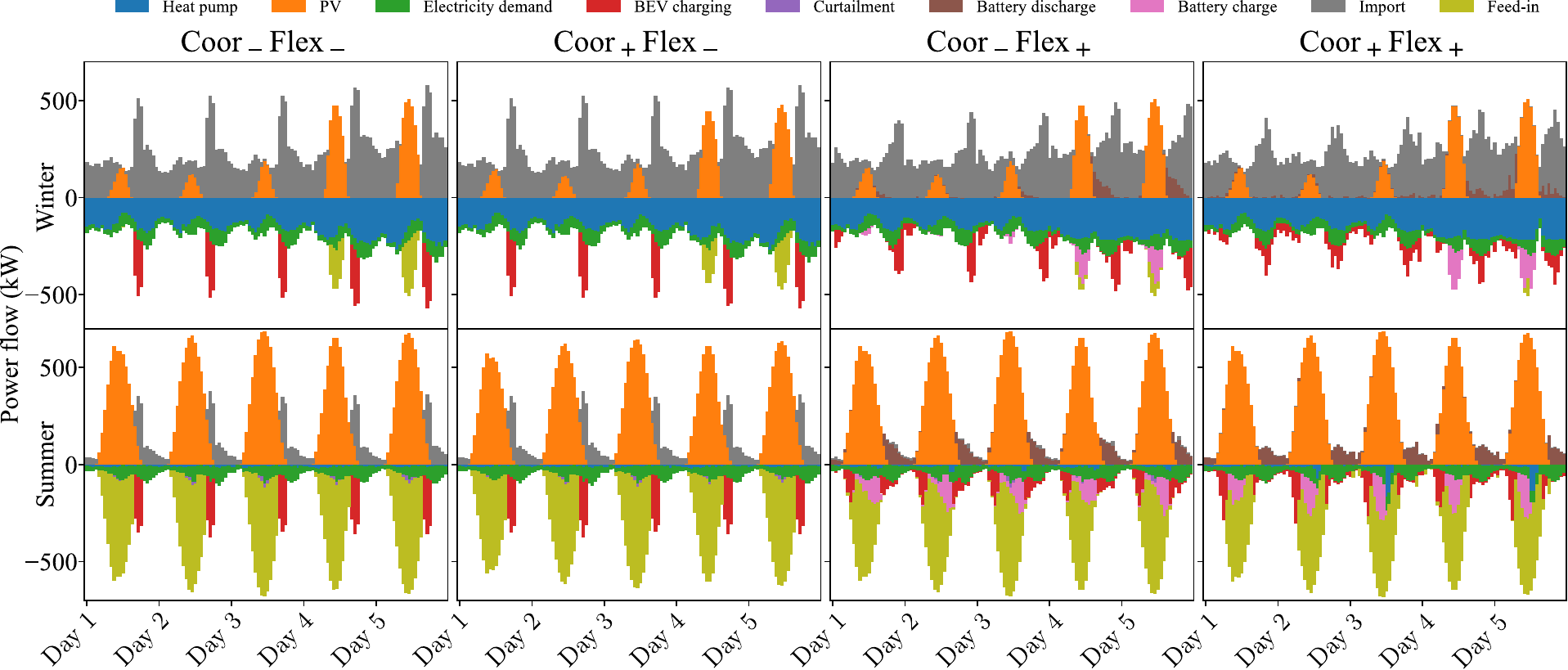}
	\caption{Electricity balance of the total distribution system in two five-day periods representing summer and winter seasons.}
	\label{fig:summerwinter}
\end{figure*}
\subsubsection{Grid reinforcement I: transformer replacement}
\begin{figure}[!ht]
	\centering	\includegraphics[width=1\linewidth]{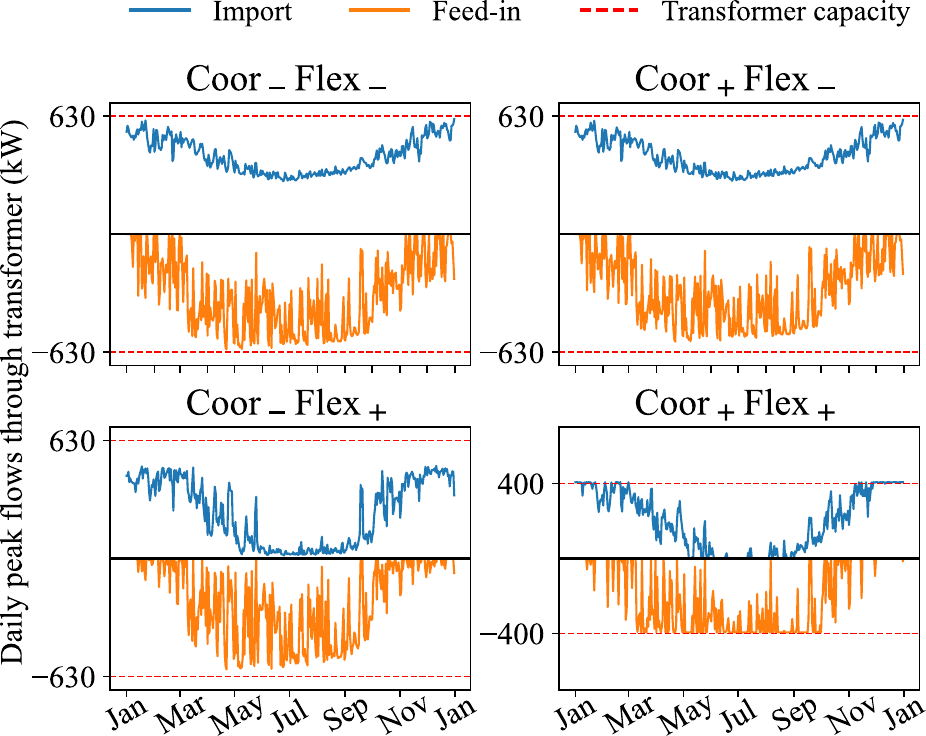}
	\caption{Daily peak import and feed-in flows through transformer over year in each paradigm, resulting from the non-convex power flow simulation.}
	\label{fig:trafo_flows}
\end{figure}

Resulting of the system behavior mentioned above, Figure \ref{fig:trafo_flows} illustrates the daily peaks of the import and feed-in flows passing through the transformer over the entire year, along with the optimized transformer capacities (dashed red lines). In all paradigms aside from \textit{Coor$_+$Flex$_+$}, the electrification of demands in winter and the feed-in of overproduced PV electricity necessitates an increase in the transformer capacity to 630 kVA compared to the pre-installed 400 kVA. The synergy between coordination and flexibilities is also visible here---an adequate peak-reducing behavior on the grid scale can only be orchestrated when the flexibility options are present in the system. Correspondingly, in the \textit{Coor$_+$Flex$_+$} paradigm, the peaky charging and heating demands can be flattened. This, combined with a more localized utilization of PV electricity, allows doing away with the need to replace the transformer altogether. Figure \ref{fig:trafo_flows} also illustrates the modest significance of the power flow approximation errors from the LinDistFlow model. Indeed, in the \textit{Coor$_+$Flex$_+$} paradigm, where the pre-installed transformer is operated at its defined limit of 400 kVA during the optimization step, occasional transformer flows exceeding the rated capacity by up to 8 kVA (2\%) are observed as the result of the losses calculated by the non-convex power flow simulation. Nevertheless, these loss-compensating excesses occur no more than $\sim$100 hours of the year, and most transformers permit such moderate, short-term overloads above their rated capacities \cite{OverloadGuide2012}.  
\subsubsection{Grid reinforcement II: parallel cabling}
\begin{figure*}[!t]
	\centering	\includegraphics[width=1\linewidth]{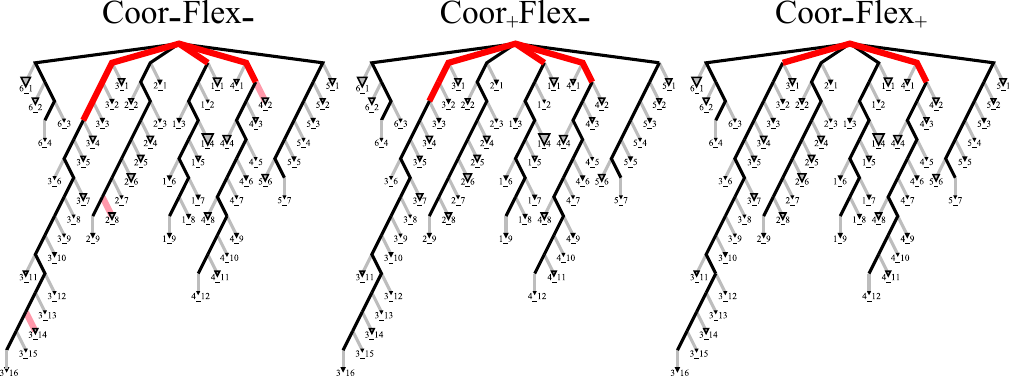}
	\caption{Optimized parallel cable reinforcement in each paradigm. The reinforced cable sections are illustrated in red, non-reinforced main line sections in black, and non-reinforced service line sections in gray. Building nodes are sized corresponding to their demands. No cable reinforcement takes place in \textit{Coor$_+$Flex$_{+}$}; therefore, this paradigm is omitted in this figure.}
	\label{fig:cable_build}
\end{figure*}

\begin{figure*}[!t]
	\centering
	\includegraphics[width=1\linewidth]{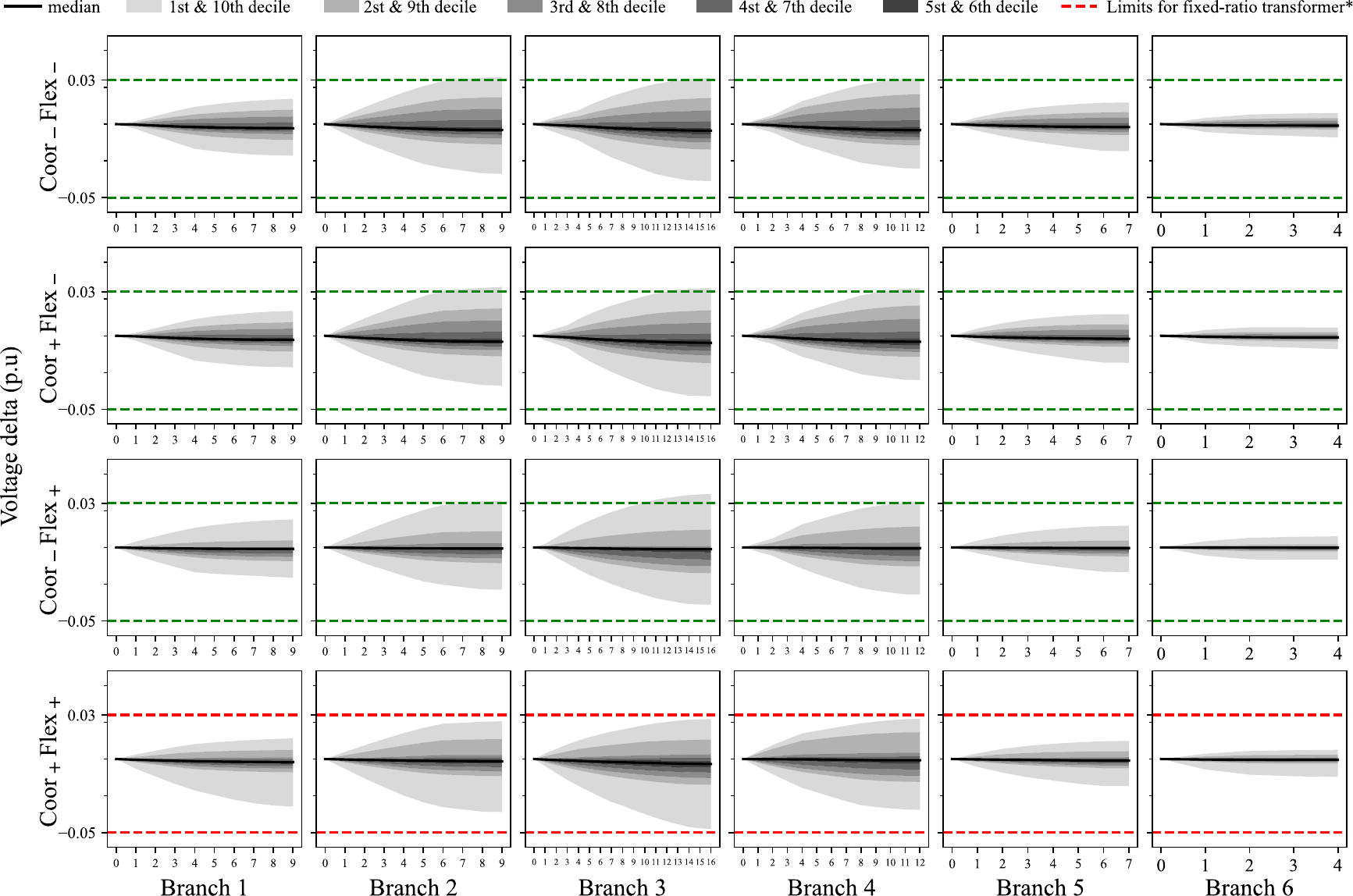}
	\caption{Deviations of nodal voltages from the main busbar in each branch for various paradigms, resulting from the non-convex power flow simulation. The paradigms are listed vertically, and each grid branch horizontally. In each plot, the x-axis represents the nodes located in the branch. Black lines represent the average voltage deviation for each node throughout the year, and the gray areas with different tones illustrate the ranges for various deciles. *The green dashed lines denote the replacement of fixed-ratio transformer with an OLTC, thereby relaxing the voltage limitation to $\pm10\%$ p.u.}
	\label{fig:voltages}
\end{figure*}

Figure \ref{fig:cable_build} shows the grid layout for the paradigms where cable reinforcement occurred, with each reinforced line section drawn in bold red. In the coordinated \textit{Coor$_+$Flex$_+$} paradigm, a result similar to the transformer replacement is observed---also here, power flows within the grid can be managed in a manner that respects all limits of existing cables, thereby avoiding their reinforcement. On the contrary, reinforcements with a second cable were needed in a varying number of line segments for the remaining paradigms. Two observations are as follows:
\begin{itemize}
	\item as the line sections that lie on the longer branches and are closer to the transformer carry the highest loads overall, they usually needed to be reinforced (see branches 1, 3, and 4), and
 	\item both coordination and flexibilities negate the necessity to reinforce service lines and reduce  that of the main lines---coordinated planning achieves this by the smaller sizing of the PV capacities, and the flexibilities through better integration of PV and flattened demands. 
\end{itemize}
In order to identify the drivers of the cable reinforcement, an analysis of the bus voltages is made. Figure \ref{fig:voltages} shows the spreads of the voltage deviations relative to the transformer at each node for each grid branch, with the distributions depicted in decile resolution. For each node, the different values in the spread represent the variations in its voltage levels throughout all time steps. The dashed lines represent the voltage restrictions with FRTs according to the norms. These limits, however, apply only to the \textit{Coor$_+$Flex$_+$} paradigm. In the other paradigms, the FRTs are replaced by the OLTCs, which allow a larger band ($\pm10\%$) as per (\ref{eq:voltagelimits}).

In almost every moment of the yearly operation, the nodal voltages stay within the limits of an FRT---in the case of \textit{Coor$_+$Flex$_+$}, no violations exist at all. Hence, the transformer exchange is not necessary for this paradigm. In the other paradigms, rare overvoltages over $+3\%$ occur on branches 2, 3, and 4. In these cases, voltage regulation by the installed OLTC becomes necessary. In order to identify whether the voltage or the loading limits drive the cable reinforcements in the uncoordinated paradigms, the following observation is made. If one "reverts" all the cable reinforcements that took place, which corresponds to halving all the impedances, this ultimately has the effect of scaling all the voltage deviations by a factor of $\sqrt{2}$\footnote{As per (\ref{eqn:lindistflow}), the deviations in squared voltages correlate linearly with the impedance so that the deviations in voltages correlate with a factor of $\frac{1}{\sqrt 2}$.}. Even in this case, the voltages would still stay within the allowable range of an OLTC. Therefore, one can conclude that the cables are reinforced due to overloading in these paradigms. 
\subsubsection{Cost and benefit analysis}
In this subsection, a comparison of the annualized system costs across the paradigms along with the costs avoided through grid reinforcement is made. Table \ref{tab: systemcosts} gives an overview of the annualized costs. For clarity, significant changes in cost components between paradigms are denoted with $\downarrow$ or $\uparrow$.

\begin{table}[!htp]\centering
	\caption{Cost breakdown for each paradigm.}\label{tab: systemcosts}
	\scriptsize
       \setlength{\tabcolsep}{5pt}
	\begin{tabular}{lrrrrr}
		&Costs (\euro/a) &Coor$_{-}$Flex$_{-}$ &Coor$_{+}$Flex$_{-}$ &Coor$_{-}$Flex$_{+}$ &Coor$_{+}$Flex$_{+}$ \\\midrule
		\multirow{8}{*}{\rotatebox[origin=c]{90}{\centering {Buildings}}} & Import & 446,684 & 437,447 & $\downarrow$ 299,400 & 285,620 \\
        & Feed-in & -52,525 & -46,447 & $\uparrow$ -28,836 & -26,718 \\
        & PV & 88,584 & 81,322 & 88,584 & 88,584 \\
        & Heat pump & 93,549 & 93,549 & $\downarrow$ 57,402 & 56,589 \\
        & Battery & 0 & 0 & $\uparrow$ 26,009 & 22,293 \\
        & Th. storage & 0 & 0 & $\uparrow$ 8,807 & 8,822 \\
        & \textbf{Total} & \textbf{576,292}& \textbf{565,871} & \textbf{451,368} & \textbf{435,189} \\
        & \textbf{Percentage} & \textbf{100.00\%} & \textbf{98.19\%} & \textbf{78.32\%} & \textbf{75.52\%} \\\midrule
		\multirow{7}{*}{\rotatebox[origin=c]{90}{\centering {Grid}}}  & Cables & 4,161 & $\downarrow$ 2,539 & $\downarrow$ 1,236 & $\downarrow$ 0 \\
            & OLTC & 1,499 & 1,499 & 1,499 & $\downarrow$ 0 \\
            & Curtailment & 263 & 199 & 138 & 16 \\
            & Q comp & 2,505 + 272* & 2,444 + 269* & 2,505 + 176* & 2,437 + 196* \\
            & P$_\text{loss}$ comp & 935* & 934* & 607* & 684* \\
            & \textbf{Total} & \textbf{9,636} & \textbf{7,884} & \textbf{6,162} & \textbf{3,333} \\& \textbf{Percentage} & \textbf{100.00\%} & \textbf{81.82\%} & \textbf{63.95\%} & \textbf{34.59\%} \\\midrule
		&\textbf{Total costs} &\textbf{585,928} & \textbf{573,755} & \textbf{457,530} & \textbf{438,522} \\
		&\textbf{Share grid} &\textbf{1.67\%} &\textbf{1.39\%} &\textbf{1.37\%} &\textbf{0.77\%} \\
		&\textbf{Percentage} &\textbf{100.0\%} &\textbf{97.92\%} &\textbf{78.09\%} &\textbf{74.84\%} \\\midrule
        \multicolumn{6}{l}{\makecell[lc]{*The cost terms given in asterisks correspond to the compensation of the losses \\ as derived from the post-optimization load flow simulation.}}
	\end{tabular}
 \end{table}

The system costs are divided into i) the costs associated with the buildings and ii) the grid-side costs. While the building component costs experience only a minor improvement through coordination (-1.8\%), flexibilities offer a greater benefit with -21.7\% (an observation similar to \cite{MatthiasHuber.2013}). This reduction comes mainly from the better integration of the PV electricity with batteries. Furthermore, heat pumps can be dimensioned with capacities smaller than the peak heating demands. These peaks can be covered instead by discharging thermal storages that had been charged over multiple hours in the preceding time steps. The synergy in combining coordination and flexibilities is evident also here: the incremental cost improvement by coordinated planning with flexibilities is higher (2.8\%) than without them (1.8\%) as imports get further reduced.

The next section of Table \ref{tab: systemcosts} gives a breakdown of the grid costs. These are associated with the grid reinforcement, the payments for the curtailed energy, the reactive power supply, and the procurement for compensating network losses\footnote{For the compensation of grid losses, an average price of 5.43\euro/MWh according to the German Federal Network Agency is assumed \cite{GermanFederalNetworkAgency.Losses}.}. In the \textit{Coor$_-$Flex$_{-}$} paradigm, the grid reinforcement costs make up the majority of the grid costs. On the other hand, in line with the operational results discussed earlier, coordination and flexibilities lead to significant reductions in the grid costs (18\% and 36\% respectively). Complete avoidance of grid reinforcement in \textit{Coor$_+$Flex$_{+}$} reduces the overall grid costs by up to 65\%, leaving only to account for the costs associated with the reactive power supply and compensation for the losses. Thanks to the optimization approach, the occasional curtailment of PV electricity by the grid operator is utilized as a supplementary, low-cost active measure that alleviates the grid reinforcement needs along with the passive measures. 

Across all paradigms, the grid-side costs make up only a minor share (less than 2\%) of the total system costs. Yet, this observation does not underplay the importance of grid reinforcement. In order to quantify the value unlocked by grid reinforcement, an analysis regarding the facilitated flows is made. Each paradigm is optimized again with fixed grid capacities and a relaxed power balance to prevent infeasibility. The resultant load shedding is then interpreted as the energy flows which were made possible through the expansion of the grid capacities. In order to calculate the corresponding avoided costs, the unmet load is priced with a domestic value of lost load (VoLL) of 13\euro/kWh for Germany \cite{CambridgeEconomicPolicyAssociatesLtd.2018}. Table \ref{tab: avoided_costs} illustrates the resultant avoided costs. For all paradigms where grid reinforcement was necessary, the costs avoided by demand satisfaction are striking: without the expansion of grid capacities, up to 3\% of the import demand would need to be shed, which nonetheless corresponds to substantial welfare losses of up to 400,000\euro. For urban distribution systems with denser localization of demands, these load-shedding proportions are expected to be even larger. The social value of demand satisfaction significantly outweighs the moderate reinforcement costs. Reflecting also on the fact that grid infrastructure projects are often associated with long lead times, the presented results underline the need to consider these investments in a timely and systematic manner, especially in high-density regions.

\begin{table}[!htp]\centering
	\caption{Avoided costs of unmet demand through grid reinforcement.}\label{tab: avoided_costs}
	\scriptsize
       \setlength{\tabcolsep}{5pt}
	\begin{tabular}{rrrrc}
		Facilitated flows &Coor$_{-}$Flex$_{-}$ &Coor$_{+}$Flex$_{-}$ &Coor$_{-}$Flex$_{+}$ &Coor$_{+}$Flex$_{+}$ \\\midrule
		 Supply (kWh) & 32,102 & 32,098 & 6,886 & \textit{no} \\
              Share in imports & 3.01\% & 2.99\% & 0.96\% & \textit{grid} \\
             \textbf{Cost (\euro/a)} & \textbf{417,329} & \textbf{417,277}&\textbf{89,521} & \textit{reinforcement} \\\midrule
	\end{tabular}	
\end{table}
\section{Conclusion} \label{sec:conclusion}
This work introduced a novel framework for multi-modal energy system optimization at the distribution grid level that co-optimizes both building and grid components, with a relatively higher focus on the distribution grid reinforcement measures. The tractability of the model formulation was put into demonstration via a case study, where a variety of planning paradigms was simulated. Through these variations, the economic potentials of flexibility and coordination aspects, covering a wide range of paradigms from the status-quo to an idealized state of a distribution system, were quantified.

The case study results confirm the significant economic value of flexibilities in a system with high PV, HP and BEV penetration, which can be already achieved with commercially available HEMS capabilities. Furthermore, they show that the total system costs deviate only minorly when system-wide coordination is facilitated. Nevertheless, coordination in the form of local energy exchange does further alleviate system peaks, especially when flexibilities are also present in the system, reducing grid reinforcement requirements significantly. Therefore, it could be considered as a supplementary measure, especially in settings where the implementation of grid infrastructure projects is less practical. The avoided cost analysis shows, on the other hand, the social value generated by expanded grid capacities to be substantial, pronouncing the importance of the eventual grid reinforcement for reliable demand satisfaction of consumers. The cost breakdown demonstrates the relative subordinance of the grid costs compared to the costs tied to investing in the building components and electricity import. However, this statement comes with two caveats---First, the cost assumptions associated with grid reinforcements represent, at best, the status-quo. However, increasing demand for grid reinforcements due to rapid uptake of electrification may lead to shortages in material and personnel, leading to higher costs or even strict limitations in expansion. Second, spendings on grid reinforcement are in reality reflected by the grid planner to the consumers as "grid surcharges" on the retail price. This constitutes a dynamic interaction between the grid operator and the consumers, leading to higher long-term retail prices for the consumers in the paradigms with high grid costs. This interaction is, however, not considered in this study.

Regardless, the minor system improvement observed through the coordinated optimization may imply that the overall welfare loss from optimizing individual buildings, instead of making holistic planning decisions with system interactions, is relatively low. Therefore, in the face of computational restrictions, a communal planner may opt to prioritize a sequential optimization of highly detailed building models with a subsequent grid planning (as in \textit{Coor}$_{-}$\textit{Flex}$_{+}${}), over holistic models (as in \textit{Coor}$_{+}$\textit{Flex}$_{+}$) that necessitate the mentioned simplifications. Note, however, that this statement holds primarily for the investigated case study of a heat supply exclusively with heat pumps. These are individual heating systems and allow decoupling of heat supply between buildings. In contrast, the possibility of a district heating alternative introduces inter-agent coupling beyond the electrical grid and hence does increase the necessity for a holistic perspective.

\subsection{Outlook}
The focus of this study was on the HOODS framework development and its application to an exemplary distribution system to demonstrate the analysis possibilities it can facilitate. A multitude of further research can exist on top of the presented framework. First, if the intended use of the framework is the holistic energy planning of a given supply region, further options for heating, i.e., pellet heaters, district heating, or solar thermal, can be introduced rather than the pre-determined component of the heat pump, as assumed in this study. Also, the retrofitting of buildings, a significant factor for the heating demand and the integration costs of heat pumps, can be introduced as an endogenous decision variable in the model. These together would enable the investigation of various fields of tension between technologies, e.g., the combination of better-insulated buildings connected to a district heating network, where a grid reinforcement is highly undesirable. Through such a variety of technologies, combined with representative grids for a variety of consumer regions (e.g., urban, suburban, or rural), the developed framework can present energy supply solutions and the accompanying grid reinforcement requirements for a wide range of distribution systems with diverse characteristics. In this process, one has to keep in mind the possibly increasing computational complexity which can be moderated with the presented time series aggregation method. 

Furthermore, a higher level of depth can be pursued for the modeling of the mobility demand. In this work, a rather simplistic, constant daily demand for each building that corresponds to the current prognoses has been assumed. By drawing on recent studies and surveys on electric mobility, detailed insights can be introduced to the model to reach a more realistic representation, such as different consumer demand patterns (e.g., weekday/weekend), the inclusion of alternative charging possibilities (e.g., charging at work, or fast charging), and the temperature dependence of the auxiliary electricity consumption in BEVs (e.g., for heating in the car). As an advanced aspect, benefits from regulations that allow vehicle-to-grid services can be assessed. Moreover, further regulations that are partially in effect, such as blockage hours or discounted electricity prices for operating heat pumps, can be introduced to the model. The mentioned grid surcharge dynamics between the consumers and the grid operator can also be implemented within a multi-level or possibly iterative model. For instance, various capacity pricing schemes for consumers can be implemented with ease, thereby investigating possible variations on the existing network tariff design that distributes grid costs on a cost-by-cause basis.

The tractability of the model framework was demonstrated on an LV grid with a representative size. As an extension, it can also be applied on higher voltage grids, such as medium-voltage grids that span wider regions, where higher grid costs are expected. In this, tradeoffs between computability and the granular modeling of building components have to be carefully considered. If necessary, the time-series aggregation can be supplemented with a spatial clustering to alleviate the computational complexity.
Additionally, the adopted deterministic approach for grid planning may be enhanced with robust optimization methods that ensure fail-safe operation over a broad set of load scenarios. Finally, the solution quality of the grid planning strategies obtained from the adopted methodology can be compared to existing heuristic methods within a comparative study.

\section*{Acknowledgments}
The authors thank the Bavarian Research Foundation ("Bayerische Forschungsstiftung") as the presented work is funded by them through the research project STROM. The authors also thank Leonhard Odersky (Technical University of Munich) for their feedback on the initial manuscript.
\section*{Declaration of competing interests}
Authors have no competing interests to declare.
\section*{Credit author statement}
\textbf{Soner Candas:} Conceptualization, Methodology, Investigation, Data curation, Writing - Original draft preparation
\textbf{Beneharo Reveron Baecker:} Methodology, Data curation, Writing - Review \& Editing 
\textbf{Anurag Mohapatra:} Writing - Review \& Editing 
\textbf{Thomas Hamacher:} Conceptualization, Funding acquisition, Writing - Review \& Editing 
\bibliographystyle{elsart-num}
\newpage
\begingroup
\bibliography{bib/biblio}
\endgroup

\newpage
{\appendices
	\section{Linearized branch flow limits}\label{app: pq}
	\begin{figure}[!htb]
		\centering
		\includegraphics[width=2.5in]{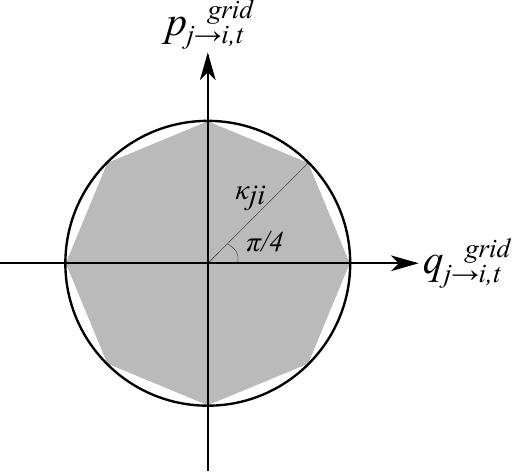}
		\caption{Linearized feasible space given a thermal line limit of $\kappa_{ji}$ and angle steps $\pi/4$.}
		\label{fig:thermal_limit}
	\end{figure}
	
	\begin{table}[!htb]\footnotesize
		\caption{Coefficients for constraint (\ref{eqn:line_lim}).}
		\centering
		\begin{tabular}{cccc}
			$y$ & $a_y$& $b_y$& $c_y$\\
			\hline
			$1$& $\phantom{-}1$ & $\phantom{-}\sqrt{2}+1$ & $\sqrt{2}+1$ \\
			$2$& $\phantom{-}1$ & $\phantom{-}\sqrt{2}-1$ & $\phantom{\sqrt{2}+{}}1$ \\
			$3$& $\phantom{-}1$ & $-\sqrt{2}+1$ & $\phantom{\sqrt{2}+{}}1$ \\
			$4$&$-1$ & $\phantom{-}\sqrt{2}+1$ & $\sqrt{2}+1$ \\
			$5$&$-1$ & $-\sqrt{2}-1$ & $\sqrt{2}+1$ \\
			$6$&$-1$ & $-\sqrt{2}+1$ & $\phantom{\sqrt{2}+{}}1$ \\
			$7$&$-1$ & $\phantom{-}\sqrt{2}-1$ & $\phantom{\sqrt{2}+{}}1$ \\
			$8$& $\phantom{-}1$ & $-\sqrt{2}-1$ & $\sqrt{2}+1$ \\		
		\end{tabular}
	\end{table}
	\section{Accuracy of time-series aggregation}\label{app: tsam}
	
	{\centering
		\includegraphics[width=2.5in]{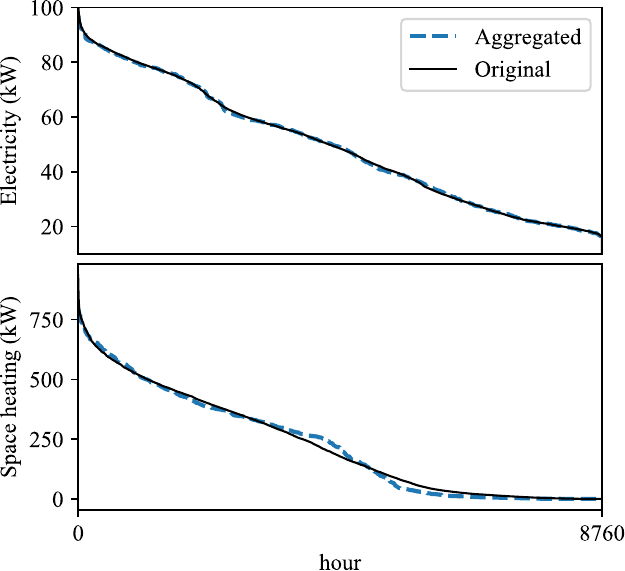}
		\captionof{figure}{Ordered duration curves for the total electricity and space heating demand in the grid.}
		\label{fig:tsam_accuracy}}
		\vfill\eject
		\section{Algorithms of the planning paradigms}\label{app: alg}
		\begin{algorithm}[H]
			\caption{Coordinated planning paradigm, with flexibilities (\textit{Coor$_+$Flex$_+$}).}
			\begin{algorithmic}
				\STATE \textbf{solve }{\textsc{HOODS($\mathcal{T}_r$}, flexible=\textbf{True})},\textbf{ return }optimal $\boldsymbol \kappa^*,\boldsymbol \alpha^*\nonumber$
				\STATE \textbf{solve }{\textsc{HOODS($\mathcal{T}$}, flexible=\textbf{True})} with $\boldsymbol \kappa = \boldsymbol \kappa^*, \boldsymbol \alpha = \boldsymbol \alpha^*$
			\end{algorithmic}
			\label{alg1}
		\end{algorithm}
  		\begin{algorithm}[H]
			\caption{Coordinated planning paradigm, without flexibilities (\textit{Coor$_+$Flex$_-$}).}
			\begin{algorithmic}
				\STATE \textbf{solve }{\textsc{HOODS($\mathcal{T}_r$}, flexible=\textbf{False})},\textbf{ return }optimal $\boldsymbol \kappa^*,\boldsymbol \alpha^*\nonumber$
				\STATE \textbf{solve }{\textsc{HOODS($\mathcal{T}$}, flexible=\textbf{False})} with $\boldsymbol \kappa = \boldsymbol \kappa^*, \boldsymbol \alpha = \boldsymbol \alpha^*$
			\end{algorithmic}
			\label{alg4}
		\end{algorithm}
		\begin{algorithm}[H]
			\caption{Uncoordinated planning paradigm, without flexibilities (\textit{Coor$_-$Flex$_-$}).}
			\begin{algorithmic}
				\STATE For all  $i\in\mathcal{I}^b$:
				\STATE \hspace{0.5cm} \textbf{solve }\textsc{HOODS-Bui}($\mathcal{T}_r,i,$ flexible=\textbf{False}), \STATE \hspace{0.5cm} \hspace{1cm}\textbf{return } \text{optimized }$\boldsymbol \kappa_i^*$ 
				\STATE  \hspace{0.5cm} \textbf{solve }\textsc{HOODS-Bui}($\mathcal{T},i, $ flexible=\textbf{False}) with $\boldsymbol \kappa_{i} = \boldsymbol \kappa_{i}^*$ 
				\STATE \hspace{0.5cm} \hspace{1cm}\textbf{return }$\text{optimized }\boldsymbol{p}_{i}^{{\text{imp}},*},  \boldsymbol{p}_{i}^{{\text{feed-in}},*}, \boldsymbol{q}_{i}^{{\text{imp}},*}$\\
				\hspace{-0.2cm} \textbf{solve }{\textsc{HOODS-Grid}($\mathcal{T},\boldsymbol{p}_{i}^{{\text{imp}},*},  \boldsymbol{p}_{i}^{{\text{feed-in}},*}, \boldsymbol{q}_{i}^{{\text{imp}},*},$ curt=\textbf{False})}
			\end{algorithmic}
			\label{alg2}
		\end{algorithm}	
		
		\begin{algorithm}[H]
			\caption{Uncoordinated planning paradigm, with flexibilities (\textit{Coor$_-$Flex$_+$}).}
			\begin{algorithmic}
				\STATE For all  $i\in\mathcal{I}^b$:
				\STATE \hspace{0.5cm} \textbf{solve }\textsc{HOODS-Bui}($\mathcal{T}_r,i,$ flexible=\textbf{True}), \STATE \hspace{0.5cm} \hspace{1cm}\textbf{return } \text{optimized }$\boldsymbol \kappa_i^*$ 
				\STATE  \hspace{0.5cm} \textbf{solve }\textsc{HOODS-Bui}($\mathcal{T},i, $ flexible=\textbf{True}) with $\boldsymbol \kappa_{i} = \boldsymbol \kappa_{i}^*$ 
				\STATE \hspace{0.5cm} \hspace{1cm}\textbf{return }$\text{optimized }\boldsymbol{p}_{i}^{{\text{imp}},*},  \boldsymbol{p}_{i}^{{\text{feed-in}},*}, \boldsymbol{q}_{i}^{{\text{imp}},*}$\\
				\hspace{-0.2cm} \textbf{solve }{\textsc{HOODS-Grid}($\mathcal{T},\boldsymbol{p}_{i}^{{\text{imp}},*},  \boldsymbol{p}_{i}^{{\text{feed-in}},*}, \boldsymbol{q}_{i}^{{\text{imp}},*},$ curt=\textbf{False})}
			\end{algorithmic}
			\label{alg3}
		\end{algorithm}	

		\onecolumn
		\section{Model parameters}
		
		\begin{table*}[!htbp]\footnotesize
			
			\caption{Select model parameters for the case study.}
			{\centering
				\bgroup
				\begin{tabular}{cclrll}
					& &\textbf{Parameters} & \textbf{Value} & \textbf{Unit} & \textbf{Source}  \\ 
					\cline{1-6} \multirow{31}{*}{\rotatebox[origin=c]{90}{\centering \textbf{Households}}} 
					& \multirow{7}{*}{\rotatebox[origin=c]{90}{\centering PV}} 		     & Investment costs, fixed    & 4074	   & \euro 			   		       & \cite{MaxPeters.2022}\\
					& 																     & Investment costs, variable & 914 	   & \euro/kW 			   		   & \cite{MaxPeters.2022}\\ 
					&																     & O\&M costs 				  & 12.5       & \euro/kW/a 		   		   & \cite{MaxPeters.2022} \\ 
					& 														    	     & WACC					      & 0.02 	   & a$^{-1}$ 			   		   & \cite{BjarneSteffen.2020} \\ 				
					& 															         & Lifetime 		          & 20   	   & a 				   		       & \cite{MaxPeters.2022} \\ 								
					& 																     & Minimum PF 				  & $\pm$0.95  & a 				   		       & $a$  \\						
					& 																     & Full load hours 			  & 1147       & h/a				  		   & \cite{StefanPfenninger.2016} \\ \cline{2-6}
					& \multirow{6}{*}{\rotatebox[origin=c]{90}{\centering Heat pump}}    & Investment costs, fixed    & 5924 	   & \euro               		   & \cite{MaxPeters.2022}\\
					&																     & Investment costs, variable & 1440 	   & \euro/kW$_\text{el}$ 		   & \cite{MaxPeters.2022}\\ 		
					& 																     & O\&M costs 				  & 60 		   & \euro/kW$_\text{el}\cdot$ a   & \cite{MaxPeters.2022}\\ 
					& 																     & WACC 					  & 0.02 	   & a$^{-1}$ 					   & \cite{QuintelIntelligence.} \\ 				
					& 																     & Lifetime 				  & 18 		   & a 						       & \cite{MaxPeters.2022}\\ 								
					& 																     & Average COP 				  & 3.5 	   & -							   & \cite{HansMartinHenning.2013}$^b$\\ \cline{2-6}			
					& \multirow{6}{*}{\rotatebox[origin=c]{90}{\centering Battery}}      & Investment costs, content  & 1000 	   & \euro/kWh$_\text{el}$ 	       & \cite{greenAkku.de.}\\ 
					&																     & O\&M costs 			 	  & 1$\%$ 		   & of Capex/a  & \cite{SteffenFattler.2019}\\ 
					&															         & WACC 					  & 0.02	   & a$^{-1}$ 					   & \cite{QuintelIntelligence.} \\ 				
					& 																  	 & Roundtrip efficiency 	  & 0.96 	   & -  						   & \cite{SteffenFattler.2019}\\ 				
					& 																  	 & Lifetime 				  & 20 		   & a 						       & \cite{SteffenFattler.2019}\\ 								
					& 																  	 & Energy-to-power ratio 	  & 3 		   & -						       & $c$\\ \cline{2-6}			
					& \multirow{5}{*}{\rotatebox[origin=c]{90}{\centering Heat storage}} & Investment costs, content  & 194 	   & \euro/kWh$_\text{th}$         & $d$\\ 			
					& 																	 & Investment costs, power 	  & 2.5		   & \euro/kW$_\text{th}$ 		   & \cite{DanishEnergyAgency.2018}\\ 
					& 																	 & WACC 					  & 0.02 	   & a$^{-1}$  					   & \cite{QuintelIntelligence.} \\ 				
					& 																	 & Roundtrip efficiency		  & 1.00	   & -  						   & \cite{DanishEnergyAgency.2018}\\ 				
					& 																     & Lifetime 			      & 30 		   & a 							   & \cite{DanishEnergyAgency.2018}\\ \cline{2-6}			
					& \multirow{3}{*}{\rotatebox[origin=c]{90}{\centering BEV}}  	 	 & Charging rate limit 	 	  & 11 		   & kW/BEV 		   & \cite{GermanFederalMinistryfordigitalandtransport.2022}\\ 
					&															         & Charging efficiency 		  & 1.00 	   & - 							   & $c$ \\
					&																     & Daily demand 			  & 5.15 	   & kWh/day 					   & \cite{Prognos.2021,OkoInstitut.2014}\\ \cline{2-6}			
					&\multirow{2}{*}{\rotatebox[origin=c]{90}{\centering P}}    													 & Electricity price 		  & 0.45 	   & \euro/kWh$_\text{el}$ 		   &  $e$ \\ 
					&																	 & Feed-in tariff 			  & 0.06 	   & \euro/kWh$_\text{el}$ 		   & \cite{GermanBundestag.2022}\\  \cline{2-6}
					&\multirow{2}{*}{\rotatebox[origin=c]{90}{\centering Q}}  																	 & Q compensation price 		  & 0.045 	   & \euro/kVARh 		   &  $f$ \\ 
					&  																	 & Load Q/P ratio & 0.25 	   & - 		   &  \cite{JorgUweScheffler.13.11.2002}$^g$ \\ 					
				\end{tabular}
				\quad
				\begin{tabular}{cclrll}\footnotesize
					& &\textbf{Parameters} & \textbf{Value} & \textbf{Unit} & \textbf{Source}  \\ \cline{1-6}
					\multirow{31}{*}{\rotatebox[origin=c]{90}{\centering \textbf{Grid}}} 
					& \multirow{12}{*}{\rotatebox[origin=c]{90}{\centering OLTC}} & Inv. costs, 160 kVA & 13000 & \euro & \cite{FlorianTobiasSamweber.2018} \\
					& & Inv. costs, 250 kVA  & 16000 & \euro & \cite{FlorianTobiasSamweber.2018}\\ 
					& & Inv. costs, 400 kVA  & 17900 & \euro &\cite{GeorgKerber.2011} \\ 
					& & Inv. costs, 630 kVA  & 19600 & \euro &\cite{GeorgKerber.2011} \\ 
					& & Inv. costs, 800 kVA  & 20800 & \euro & \cite{GeorgKerber.2011}\\ 
					& & Inv. costs, 1000 kVA  & 21900 & \euro & \cite{GeorgKerber.2011}\\ 
					& & Inv. costs, 1250 kVA  & 24300 & \euro & \cite{GeorgKerber.2011}\\ 
					& & Inv. costs, 1600 kVA  & 26600 & \euro & \cite{GeorgKerber.2011}\\ 
					& & Inv. costs, 2000 kVA  & 29500 & \euro & \cite{GeorgKerber.2011}\\ 
					& & O\&M costs & 1\% &of Capex/a & \cite{GeorgKerber.2011} \\ 
					& & WACC & 0.06 & a$^{-1}$ & \cite{E.ONGroup.2018}\\ 		
					& & Lifetime & 40 & a &  \cite{FlorianTobiasSamweber.2018}\\ \cline{2-6}
					& \multirow{19}{*}{\rotatebox[origin=c]{90}{\centering Cables}} & Installation costs & 90 & \euro/m&  \cite{GeorgKerber.2011}\\
					& & Material costs & 10 & \euro/m&  \cite{GeorgKerber.2011}\\
					& & WACC & 0.06 & a$^{-1}$  &  \cite{E.ONGroup.2018}\\ 				
					& & Lifetime & 40 & a & \cite{GeorgKerber.2011}\\ 								
					& & Reactance,  4x150     & 0.080  & $\Omega$/km & \cite{Faber.} \\ 			
					& & Resistance, 4x150     & 0.206  & $\Omega$/km & \cite{Faber.} \\
					& & Loading limit, 4x150  & 121     & kVA 		 & $h$ \\		
					& & Reactance,  4x50      & 0.102  & $\Omega$/km & \cite{Faber.}\\ 			
					& & Resistance, 4x50      & 0.387  & $\Omega$/km & \cite{Faber.} \\ 	
					& & Loading limit, 4x50   & 60     & kVA 		 & $h$ \\ 		 		
					& & Reactance,  4x35   	  & 0.105  & $\Omega$/km & \cite{Faber.}\\ 			
					& & Resistance, 4x35      & 0.524  & $\Omega$/km & \cite{Faber.} \\ 					 		
					& & Loading limit, 4x35   & 48     & kVA 	     & $h$ \\ 	
					& & Reactance,  4x25      & 0.110  & $\Omega$/km & \cite{Faber.}\\ 			
					& & Resistance, 4x25      & 0.727     & $\Omega$/km& \cite{Faber.} \\
					& & Loading limit, 4x25   & 38     & kVA	     & $h$ \\ 		 					 		
					& & Reactance,  4x16      & 0.116  & $\Omega$/km & \cite{Faber.}\\ 			
					& & Resistance, 4x16      & 1.150    & $\Omega$/km & \cite{Faber.} \\ 			
					& & Loading limit, 4x16   & 30     & kVA         & $h$ \\ 		 		 	

				\end{tabular}
				\egroup
				\label{tab: techeco}\\}
			\footnotesize$^a$ According to the VDE-AR-N 4105 norm.\\
			\footnotesize$^b$ Calculated by taking the weighted average for COP values for generating heat in temperatures $\leq60\degree$ and $\geq60\degree$, weighted according to the demand of space heating and domestic hot water at a given time.\\
			\footnotesize$^c$ Own assumption. \\	
			\footnotesize$^d$ Corresponds to the cost of a commercial product of 1800\euro\  with 400 lt volume, assuming a temperature difference of 20$\degree$.\\
			\footnotesize$^e$ Price information from Stadtwerke Munich (\url{https://www.swm.de/strom/stromtarife}) as of June 2022\\
			\footnotesize$^f$ While no tariff exists for reactive power in Germany as yet, 10\% of the electricity price is assumed for the case study.\\
			\footnotesize$^g$ Approximated from Figure 2.23 of the source.\\
			\footnotesize$^h$ Calculated using the thermal current limits according to the DIN VDE 0276-603 norm, on which a loading limit of 70\% is applied according to DIN VDE 0298-4 norm for cables with three loaded conductors.
		\end{table*}
		\onecolumn
		\section{Demand distribution in grid}
		\begin{center}
			\centering
			\includegraphics[width=0.9\linewidth]{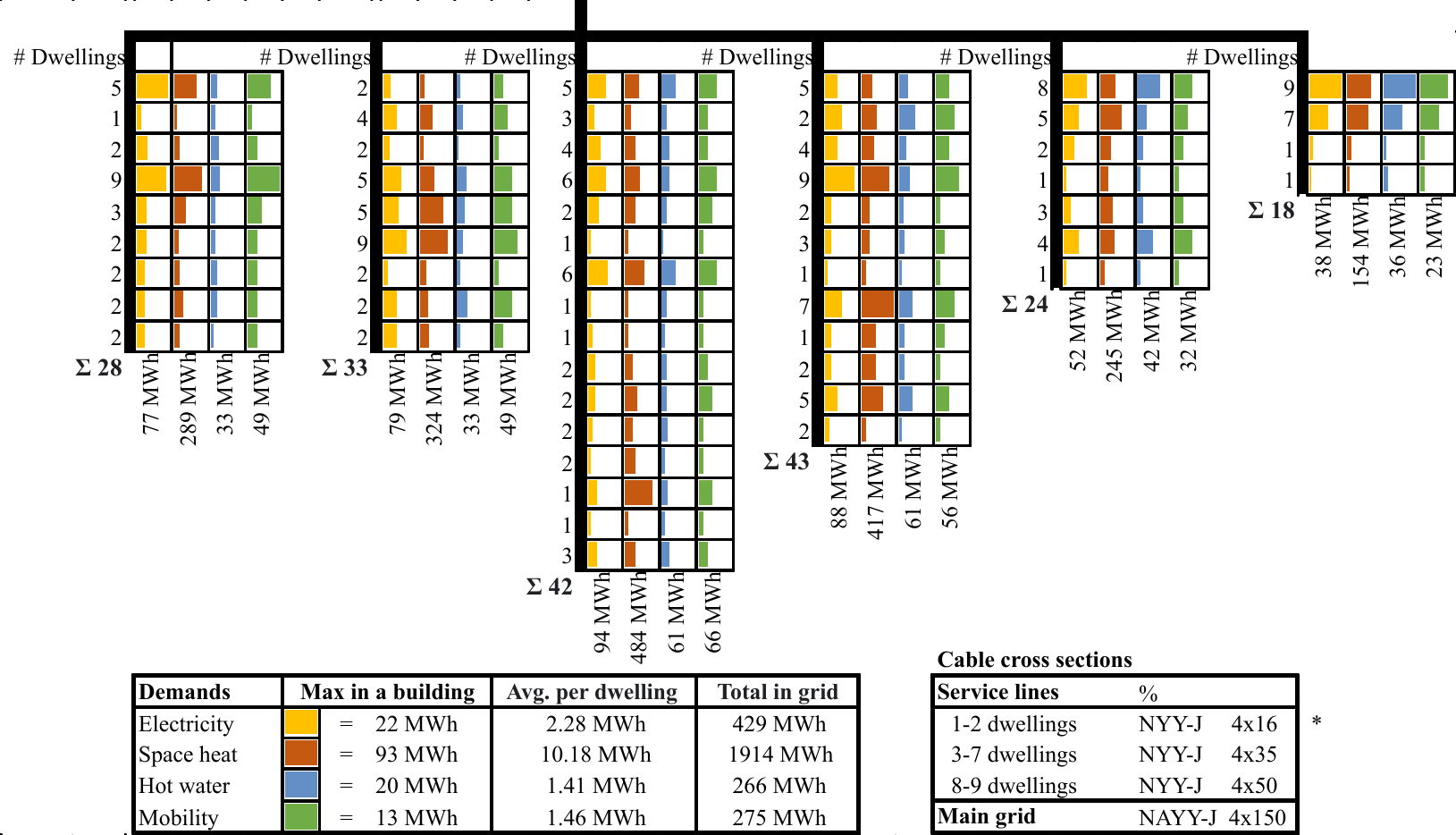}
			\captionof{figure}{Distribution of demands in each building. *Service line cross sections were determined according to the German DIN 18015-1 norm.}
			\label{fig:demand_visual}
	\end{center}

        \section{Monthly dispatch results of the distribution system}
        \begin{center}
            \centering
            \includegraphics[width=0.9\linewidth]{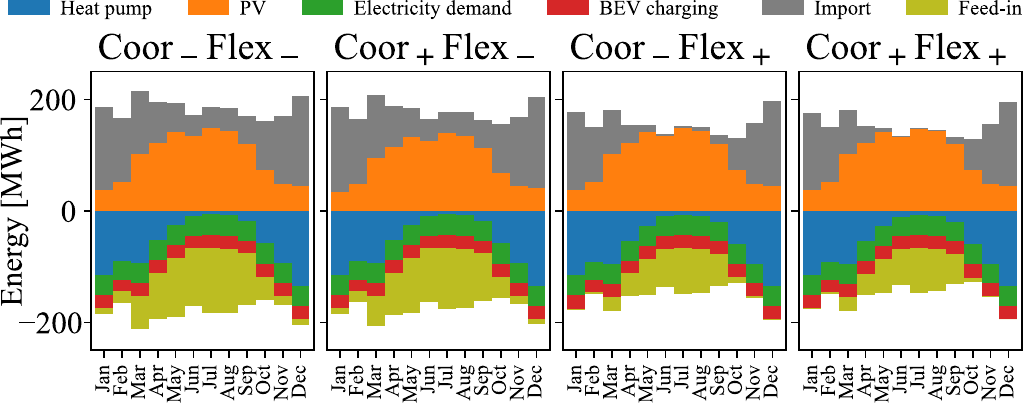}
            \captionof{figure}{Monthly sums of the electricity balance terms in the whole distribution system.}
            \label{fig:monthly_results} 
            \end{center}}
	\twocolumn

\end{document}